\pdfoutput=1

\documentclass[11pt]{article}
\usepackage{jheppub}
\usepackage{float}

\usepackage{caption}
\usepackage{subcaption}
\usepackage{extarrows}

\usepackage{mathrsfs}

\usepackage{amsmath}
\usepackage{latexsym,amssymb}
\usepackage{graphicx,color,slashed}
\usepackage{bbm} 
\usepackage{mathtools}
\usepackage{ifpdf}

\usepackage{esint}

\usepackage{amsfonts}
\usepackage{amsmath}
\usepackage{epsfig}
\usepackage{isomath}
\usepackage{latexsym,amssymb}
  \usepackage{amsmath,amssymb,amsthm}
  \DeclareSymbolFontAlphabet{\Scr}{rsfs}

\newcommand{\cF}{\mathcal{F}}

\newcommand{\cA}{\mathcal{A}}
\newcommand{\cB}{\mathcal{B}}

\newcommand{\cH}{\mathcal{H}}
\newcommand{\cL}{\mathcal{L}}
\newcommand{\cM}{\mathcal{M}}
\newcommand{\cN}{\mathcal{N}}
\newcommand{\cO}{\mathcal{O}}

\newcommand{\cV}{\mathcal{V}}
\newcommand{\Tr}{\mathrm{Tr\,}}

\newcommand{\be}{\begin{equation}}
\newcommand{\ee}{\end{equation}}
\newcommand{\ba}{\begin{eqnarray}}
\newcommand{\ea}{\end{eqnarray}}

\renewcommand{\d}{\textrm{d}}

\def\E{{$E_{7(7)}$}}
\def\EE {{$E_{5(5)}$}}
\def\EEE {{$E_{3(3)}$}}

\def\ID{\relax{\rm I\kern-.18em D}}

\renewcommand{\l}{\lambda}

\def\d{\delta}

\newcommand{\rf}[1]{(\ref{#1})}
\newcommand{\bea}{\begin{eqnarray}}
\newcommand{\eea}{\end{eqnarray}}

\def\bfzero{\relax{\rm I\kern-.18em 0}}
\def\bfone{\relax{\rm 1\kern-.35em 1}}
\def\twomat#1#2#3#4{\left(\begin{array}{cc}
\end{array}
\right)}

\def\d{\delta}

\def\cH{{\cal H}}

\newcommand{\F}{\mathds{F}}
\newcommand{\C}{\mathds{C}}

\title{\rm{\bf      The role of Sp(2n) duality in quantum theory }}
\author
{ Renata Kallosh}
 \affiliation{Stanford Institute for Theoretical Physics and Department of Physics,\\
 Stanford University, Stanford, CA 94305, USA}

\abstract{ D-dimensional maximal supergravities type I with G/H coset spaces have global G-symmetry and local H symmetry, which can be gauge-fixed in symmetric or Iwasawa-type gauges.   Maximal D-dimensional supergravities type II derived from higher dimensions without dualization have less local and global symmetries. In 4D, Gaillard-Zumino duality group Sp(56) enhances U-duality symmetry group \E\,. Using the Hamiltonian path integral, we show how extra symmetries beyond \E\, serve to establish S-matrix equivalence of supergravities I and II in different gauges. Enhanced dualities are not available in D $> 4$. This is consistent with the existence of local H symmetry and global G symmetry anomalies and UV divergences in D $> 4$ supergravities, and with the absence of these anomalies and UV divergences, so far, in 4D $\cN\geq 5$ supergravities.

}

\begin{document}

\maketitle



\parskip 4pt

\newpage

\section{Introduction}
\begin{figure}[H]
\centering
\includegraphics[scale=0.42]{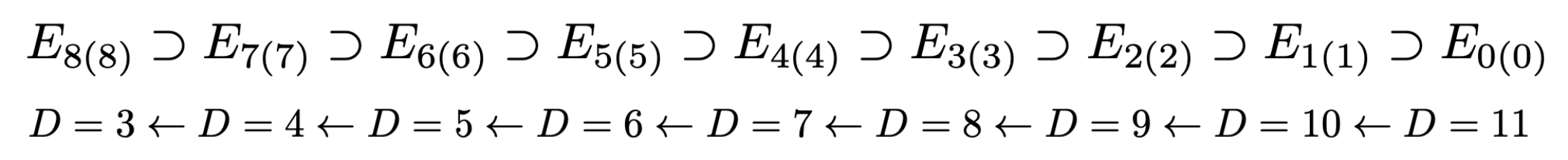}
\caption{\footnotesize  Group disintegration process proposed by B. Julia in \cite{Hawking:1981bu}, a chain of  $E_{11-D(11-D)}$  groups. These are universal maximal supergravity U-duality groups in each $D\geq 3$ dimension. 
}
\label{ScreenShot}
\end{figure}

It was shown  in \cite{Hawking:1981bu,Hull:1994ys,Andrianopoli:1996zg,Cremmer:1997ct},  using  torus  compactification of 11D or 10D supergravity and dualization, that duality symmetry  of maximal supergravities in dimension D are U-duality groups 
$  G_U= E_{11-D (11-D)}$.   These groups are often called $E_{11-D}$ \footnote{Exceptional groups $E_{11-D (11-D)}$    for $D\geq 6$ are: $E_{0(0)}$ is trivial, $E_{1(1)}= \mathbb{R}, E_{2(2)}=GL(2, \mathbb{R}), \\ E_{3(3)}=SL(3,\mathbb{R}) \times SL(2,\mathbb{R}),  E_{4(4)} = SL(5,\mathbb{R}), E_{5(5)}= SO(5,5)$.}. 
There is a successive embedding of the exceptional Lie algebras $E_{11-D}$   starting with $D=3$ all the way up to $D=11$.   It is a process introduced as a ``group disintegration'' in \cite{Hawking:1981bu}, which is inverse to dimensional reduction, see Fig. \ref{ScreenShot}.

{\it The question we would like to ask here is} : Why \E\,  symmetry appears to protect, so far,  maximal 4D supergravity from UV divergences, whereas  $E_{6(6)}$, \EE, $E_{4(4)}$, \EEE, $E_{2(2)}$ already failed to do so in all D$>$4 maximal supergravities.  There are UV divergences at some loop order, see a review of perturbative supergravity UV divergences in \cite{Bern:2023zkg}. 

\begin{figure} [H]
\centering
\includegraphics[scale=0.62]{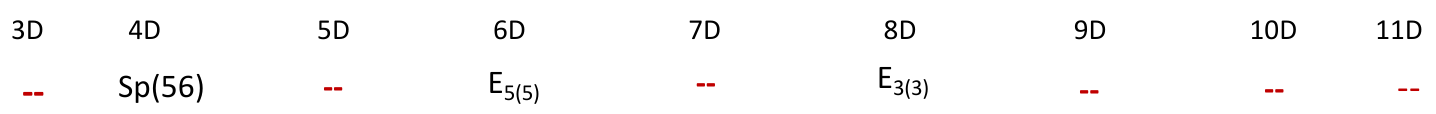}
\caption{\footnotesize  Maximal duality groups in dimension D, including Gaillard-Zumino type electro-magnetic dualities. These do not form a chain, they depend on dimensions. There are no GZ dualities in odd dimensions. Only in 4D maximal GZ duality $Sp(56)$ is a much bigger group than U-duality \E\,. In 6D and 8D, maximal duality, including GZ duality, is the same group as U-duality.
}
\label{Dualities1}
\end{figure}

The answer to this question requires taking into account electro-magnetic type duality symmetries known as  Gaillard-Zumino  (GZ) dualities \cite{Gaillard:1981rj}. We represent them in Fig. \ref{Dualities1}. These are present only in even dimensions where electric and magnetic D/2 forms can have the same dimension. They are absent in odd dimensions.

A {\it short answer} to the question above is:  In 4D   GZ duality symmetry is $Sp(56, \mathbb{R})$  \cite{Gaillard:1981rj}. It has many more symmetries than its subgroup \E\,, a U-duality group.  But in 6D and in 8D, maximal duality symmetries in Fig. \ref{Dualities1}, including GZ dualities, are exactly the same as the U-duality groups $G_U$ in Fig. \ref{ScreenShot}. These extra symmetries, which are available only in 4D and absent in D$>$4,  make 4D special. We will show how to use them in the context of a quantum theory.

A short version of these results was already presented in  \cite{Kallosh:2024ull}. Both \cite{Kallosh:2024ull} and this paper are based on the results in  \cite{Kallosh:2024rdr} where 
supergravities of type I and type II in dimension D were introduced and gauge-fixing local H symmetry in these supergravities was studied.  

Maximal supergravities in 4D, 6D, 8D were constructed in \cite{Cremmer:1979up,deWit:1982bul},  \cite{Tanii:1984zk,Bergshoeff:2007ef},  \cite{Salam:1984ft}, respectively, where it was also shown that there is a \E\, duality symmetry in 4D,  \EE\, duality symmetry in 6D and $E_{3(3)}$ symmetry in 8D.
Maximal supergravities in even dimensions 4D, 6D, 8D have Gaillard-Zumino  (GZ) \cite{Gaillard:1981rj,Tanii:1984zk,Salam:1984ft} electro-magnetic duality symmetry $Sp(56, \mathbb{R}), SO(5,5), Sp(2,  \mathbb{R})$, respectively, studied in \cite{Andrianopoli:1996ve} in detail and in \cite{Cremmer:1997ct}, in the context of doubled Lagrangians with twisted self-duality condition.  Physical scalars in these theories are coordinates of the G/H coset spaces. The groups G are $E_{11-D (11-D)}$ i. e.  \E\,, \EE \, and \EEE \,, respectively in 4D, 6D, 8D. The groups H are the maximal compact subgroups of G, and supergravities have local H symmetry, which requires gauge-fixing.

As different from local H symmetries, duality symmetries have global parameters and are characterized by duality current conservation on-shell. For example, in 4D in maximal supergravity there is a Noether-Gaillard-Zumino (NGZ) $Sp(56, \mathbb{R})$ current conservation \cite{Gaillard:1981rj, Kallosh:2008ic,Aschieri:2008ns,Hillmann:2009zf,Bossard:2010dq,Kallosh:2011dp}  which we describe shortly in Sec. \ref{Sec:NGZ}.

 Supergravity U-duality groups $G_U$ were proposed in  \cite{Hull:1994ys} as arising from the compactification of 10D supergravity on a torus and dualization.  It is the same group G in the form  $E_{d+1 (d+1)}$, where $d=10-D$. The U-duality in string theory was further suggested as a discrete subgroup of continuous dualities in supergravity.

In odd dimensions, such as 5D, 7D, 9D, the relevant G groups $E_6$, $E_4$, $E_2$ are manifest symmetries of the supergravity action. 
In even dimensions like 4D, 6D, 8D, the relevant $E_7, E_5, E_3$ symmetries are symmetries of equations of motion and Bianchi Identities. Only so-called electric subgroups of $E_7, E_5, E_3$ groups are manifest symmetries of the action.

On top of that, even dimensional supergravities have a Gaillard-Zumino  (GZ) electro-magnetic duality symmetry proposed in  \cite{Gaillard:1981rj} in 4D and in \cite{,Tanii:1984zk} in 6D and 
studied in general in \cite{Andrianopoli:1996ve} and in \cite{Cremmer:1997ct}. See also Fig. \ref{Dualities1} where it is shown that GZ dualities are dimension-dependent.

These duality groups affect scalars and form fields, vectors, and tensors. In case of $n$ such form fields, when ${D\over 2}=2p$ is even,  this duality group is  $Sp(2n, \mathbb{R})$. 
When ${D\over 2} =2p+1$  is odd this duality group is  $SO(n, n)$. Here $n$ is the number of 
($D/2-1$)-form fields, vectors in 4D, 2-tensors in 6D, 3-tensors in 8D.
Therefore for maximal supergravities, the duality group including GZ duality is
\bea
4D : \, Sp(56, \mathbb{R}) \supset E_{ 7(7)},  \quad 
6D :\,   SO(5,5)\sim E_{ 5(5)},  \quad 
8D :\,   SL(3,\mathbb{R}) \times Sp(2,\mathbb{R})\sim E_{ 3(3)}
\label{even}\eea
Only in 4D $E_{ 7(7)}$ is a subgroup of $Sp(56, \mathbb{R})$, but in 6D $E_{ 5(5)} $ is 
 the same as 
$SO(5,5)$  and in 8D $E_{ 3(3)}$ is the same as $SL(3,\mathbb{R}) \times Sp(2,\mathbb{R})$.
This means that in all maximal supergravities with D $ >4$, maximal duality, including GZ duality, has the same number of generators as a U-duality group $E_{11-D (11-D)}$ with $D> 4$.

In 4D $\cN\geq 5$
 GZ duality group is $Sp(2n_v, \mathbb{R})\supset G_U$ whereas U-duality group $G_U$ is a subgroup of $Sp(2n_v, \mathbb{R})$. In particular for $\cN\geq 5$
\bea
&&\cN=8:  \, \,   G_{GZ}=Sp(56,  \mathbb{R}) \supset E_{7(7)} \cr
&&  \cN=6:  \, \,  G_{GZ}=Sp(32,  \mathbb{R})  \supset SO^*(12)\cr
&&  \cN=5: \, \,  G_{GZ}= Sp(20,  \mathbb{R})  \supset SU(1,5)
\label{N568}\eea
This fact plays a significant role in the quantum aspects of gauge-fixed supergravities.

In each D-dimensional maximal supergravity with local H symmetry, one can consider 3 different gauges studied in detail in \cite{Kallosh:2024rdr}:
\begin{enumerate}
  \item   {\it Symmetric} local H-symmetry gauge where the physical scalars transform in a linear representation of the maximal compact subgroup $H \subset G$, and global H-invariance is present on-shell. Also, G-symmetry is preserved after gauge-fixing, but it is nonlinearly realized.

 \item {\it Iwasawa triangular gauge} of a local H-symmetry where the right node was deleted in the Dynkin diagram of $E_{11-D}$. The related D+1 theory was gauge-fixed in the Iwasawa gauge for $(G/H)_{D+1}$ before compactification on a circle of the theory from D+1 to D dimensions was performed. These gauges are related to solvable Lie subalgebras of U-dualities, see for example \cite{Trigiante:1998vu} where solvable Lie algebras are described in supergravities.
 
  \item  {\it Partial Iwasawa}  local H-symmetry gauge, where the related D+1 theory was gauge-fixed in the symmetric gauge for $(G/H)_{D+1}$ before compactification on a circle of the theory from D+1 to D dimensions was performed.
  
\end{enumerate}

An additional class of gauges considered in \cite{Kallosh:2024rdr} are triangular Iwasawa gauges related to compactification on a torus $T^{11-D}$ from 11D supergravity. Other gauges associated with $D+n$ supergravities dimensionally reduced to D-dimensions are possible.

Besides different gauges of supergravities I with a coset space $(G/H)_D$ and corresponding symmetries, we discussed in \cite{Kallosh:2024rdr} a class of D-dimensional supergravities which we call supergravities II. These are derived from (D+n) dimensions by compactification, and they have less symmetries. We studied mostly the case of n=1, i. e. D+1 supergravity compactified on a circle. Due to the group disintegration process shown in Fig. \ref{ScreenShot}, supergravities II tend to have a smaller G-symmetry. For example, in n=1 case instead of $G_{11-D (11-D)}$ they have $G_{10-D (10-D)}$. Their maximal compact subgroups, defining their local symmetries, are also smaller.

The goal of this work is to investigate the relation between different gauges of supergravities I as well as the relation between supergravities I and supergravities II.  If on-shell results from the theory in different gauges and cases I and II can be proven to agree, we may qualify it as evidence of the absence of local H-symmetry anomalies. If on-shell results from supergravities I and supergravities II agree, we can qualify this as an absence of G-symmetry anomalies. And vice versa, we will find that the absence of extra symmetries in D $>4$ above $G_U$ presents an obstacle when trying to prove the gauge-invariance and G-invariance. Therefore D $>4$ supergravities are consistent with local H-symmetry and global G-symmetry anomalies.

This analysis has to be correlated with the properties of H-symmetry/G-symmetry  anomalies and UV divergences in \cite{Kallosh:2023css,Kallosh:2023dpr,Kallosh:2024lsl} based on available amplitude computations of UV divergences. We presented some details on critical loop order and UV divergences in \cite{Kallosh:2024ull}, see also eq. (2.2) and reference [14] there. It shows that at $D>4$ all UV divergences are below  $L_{cr}$ for maximal supergravities.
 In 4D, more is known about $\cN\geq 5$ supergravities where the critical loop order is $L_{cr} = \cN$.
The loop computations tell us that so far, no UV divergences in 4D at $\cN\geq 5$  below $L_{cr}$  were found. For example, at $\cN= 5$ the 82 diagrams at $L=4 < L_{cr}=5$ cancel \cite{Bern:2014sna}. Thus, 
 at present,  D=4 is different from D $>4$  supergravities.

We will use the fact that maximal duality at D $>4$, including GZ duality in even dimensions, is the same as  the U-duality \be
G_{max}= G_U, \qquad \qquad \, \, \, \, D > 4 \ ,
\ee
whereas the  amount of symmetries in  GZ duality at 4D exceeds the   amount of symmetries in U-duality for $\cN\geq 5$
\be
G_{GZ}= Sp(2n_v) \supset G_U, \quad D=4 \ .
\ee 
Based on this, we will show that the extra symmetries in 4D  $\cN\geq 5$ supergravities due to $G_{GZ}\supset G_U$ imply that the local  H-symmetry and global G-symmetry have no anomalies.

 \section{Supergravity I and II and  abelian ideals }

Maximal supergravities in dimension D have different versions related to dimensional reduction from higher dimensions D $\leq 11$.  These versions, supergravity I and II, were defined in \cite{Kallosh:2024rdr}. The models I are the original $(G/H)_D$ supergravities in dimension D. The models II, in general, are the ones derived from D+n compactified supergravities. The case with n=1,   D+1 supergravity compactified on a circle, with inherited $(G/H)_{D+1}$ coset space, was the main focus of the studies in \cite{Kallosh:2024rdr}.

 Only supergravities I have local $H_D$-symmetry, supergravities II have a smaller local symmetry, at most a local $H_{D+1}$ symmetry associated with D+1 supergravity with its $(G/H)_{D+1}$ coset space. 4D  supergravity I is in \cite{Cremmer:1979up,deWit:1982bul}, D+1 supergravity II is in \cite{Andrianopoli:2002mf}, in 6D  supergravity I is in \cite{Tanii:1984zk,Bergshoeff:2007ef}, D+1 supergravity II is in 
\cite{Cowdall:1998rs}.
 
 The universal nature of these two supergravity versions in each dimension D follows from the fact that the duality Lie algebra has different decompositions: a Cartan decomposition and an Iwasawa decomposition.

1.  In supergravities I, all physical scalars in the action in dimension D have non-polynomial dependence in a symmetric gauge.
These are original theories constructed in \cite{Cremmer:1979up,deWit:1982bul} in 4D and in 
\cite{Tanii:1984zk,Bergshoeff:2007ef} in 6D, where the quantization in a symmetric gauge was performed in \cite{Kallosh:2024rdr}. It is associated with the {\it Cartan decomposition} of a duality symmetry  Lie algebra 
\be
\mathbb{G}= \mathbb{H}\oplus \mathbb{K}
\ee 
Here $\mathbb{H}$ is a maximal compact subalgebra of $\mathbb{G}$, and $\mathbb{K}$ includes the remaining operators associated with the coset space G/H. For example, in $\cN=8$ in 4D  there are 63 generators in   $H=SU(8)$ and 70  in ${E_{7(7)}\over SU(8)}$

2. In supergravities II,  some physical scalars in the action in dimension D have non-polynomial dependence, but some have a polynomial dependence.  They are  associated with the {\it Iwasawa decomposition} of a duality symmetry Lie algebra 
\be
\mathbb{G}= \mathbb{H}\oplus \mathscr {S}\, , \qquad \mathscr {S}={ C}\oplus { N}\,.\label{SCN}
\ee 
Here $\mathscr {S}$ is a solvable subgroup of the Lie algebra,  
${ C}$ is the Cartan subspace of the coset space and  $N$ is a nilpotent subalgebra.
The supergravity models II, which we focus on here,  originate from D+1 dimension, they were described in \cite{Andrianopoli:1996zg}.

In supergravity I, it is possible to fix the local $H_D$-symmetry in a symmetric gauge where all scalars interact non-polynimially. Alternatively,  it is possible to fix it in one of Iwasawa or partial Iwasawa gauges where some scalars, axions, enter the action non-polynomially. In supergravities II it is only possible to gauge-fix a local $H_{D+1}$-symmetry in symmetric or Iwasawa gauges, but there are always some axionic scalars in dimension D in supergravities II. Besides having axions, supergravities II have form fields, which do not form representations of the $G_D$-symmetry, as they do in supergravities II as shown in \cite{Kallosh:2024rdr}.

 In  supergravity II, the minimal {\it number of axions is given by the dimension of the maximal abelian ideal of a solvable Lie
algebra}, defined as the maximal subset of nilpotent generators commuting among themselves \cite{Andrianopoli:1996zg}.

The maximal abelian ideal in dimension D can be characterized by
 the decomposition of the U--duality algebra $E_{r+1(r+1)}$ with
respect to the U--duality algebra in dimension D+1.  Note that $r+1= 11-D$ and $r= 11-(D+1)$.

Consider the decomposition of $E_{r+1(r+1)}$ with respect to the subalgebra
\be
E_{r+1(r+1)} \to E_{r (r )} \, \otimes \, O(1,1)
\label{decomp}\ee 
 There is a general pattern in this decomposition:
\begin{equation}
 \mbox{adj }  E_{r+1(r+1)} \, = \, \mbox{adj }  E_{r(r)}  \, \oplus
 \, \mbox{adj }  O(1,1) \, \oplus ( \ID^+_{r} \oplus \ID^-_{r} )
 \label{genpat}
\end{equation}
Here $\ID^+_{r}$ is  an irreducible representation of
the  U--duality algebra $E_{r (r )}$ in $D+1$ dimensions. It  coincides with the
maximal abelian ideal ${\cal A}_{r+1}$
\begin{equation}
\ID^+_{r}    \, \equiv \, {\cal A}_{r+1}   \, \subset \, Solv_{(r+1)}
\label{genmaxab}
\end{equation}
of the solvable  Lie algebra. 
The subspace $\ID^-_{r}$ is  a second identical  of the representation
 $\ID^{+}_{r}$  made of negative rather than of positive weights
 of $E_{r (r )}$. 
Also,    $\ID^{+}_{r}$  and   $\ID^-_{r}$
 correspond to the eigenspaces belonging respectively to the eigenvalues
 $\pm 1$ with respect to  $O(1,1)$.

It was explained in \cite{Andrianopoli:1996zg} that, from physics perspective, the dimension of the abelian
ideal in $D$ dimensions is equal to the number of vectors in dimensions $D+1$.
It follows from \rf{genpat}
that the total dimension of the abelian ideal is given by
\be
{\rm dim} \, \cA_{D} \,\equiv \,  {\rm dim} \, \cA_{r+1}
 \,\equiv \,  {\rm dim} \,
\ID_{r}
\label{abeliii}
\ee
where $\ID_{r} $ is a representation of $U_{D+1}$ related to vector fields.
 According to \rf{genpat} for $D \ge 4$:
$
\mbox{adj } U_D = \mbox{adj } U_{D+1} \oplus {\bf 1} \oplus ({\bf 2}, \ID_r)
$.
This is  a consequence of the embedding chain in Fig. \ref{ScreenShot}
which at the first level of iteration yields \rf{decomp}.

For example, under $E_7 \rightarrow E_6 \times O(1,1)$ we have the branching rule:
${\rm adj} \, E_7 = {\rm adj} \, E_6 + {\bf 1} + ({\bf 2},{\bf 27})$ and
the abelian ideal is given by the ${\bf 27^+}$ representation of the $E_{6(6)}$ group.
The $70$ scalars of the 4D $\cN=8$ supergravity II  are
 decomposed as 
\be
{\bf 70} = {\bf 42} +{\bf 1} +{\bf 27^+}
\label{70}\ee
This is a property of scalars in the Iwasawa gauge.
Note that in supergravity I in the standard symmetric gauge  {\bf 70} is a representation of $SU(8)$.

In general, the cosets in D,  $G_D/H_D$,  are related to cosets in D+1,  $G_{D+1}/H_{D+1}$,  as follows
\be
\frac{G_D}{H_D} \sim \Big ( \frac{G_{D+1}}{H_{D+1}}, r_{D+1}, {\bf V}_r^{D+1}\Big)
\label{cosetideal}
\ee
Here $r_{D+1}$ is a compactification radius, and ${\bf V}_r^{D+1} $
are the compactified vectors presenting  the abelian ideal in $D$ dimensions and 
\begin{equation}
  \mbox{adj}\, H_D = \mbox{adj} \, H_{D+1} \, +\,
\mbox{adj} \,\mbox{Irrep} \, U_{D+1}
\end{equation}
for example 
${\rm adj} \, SU(8) = {\rm adj} \, USp(8) \oplus {\bf 27^-}$
$ \Longrightarrow $ ${\bf 63}= {\bf 36}+ {\bf 27^-}$. In 6D ${\rm adj} \, (SO(5)\times SO(5)) = {\rm adj} \, SO(5) \oplus {\bf 10^-}$
$ \Longrightarrow $ ${\bf 20}= {\bf 10}+ {\bf 10^-}$.

Thus, the Iwasawa type gauges in 4D and in 6D supergravities are related to 5D and 7D supergravities due to properties of the solvable Lie algebra and their maximal abelian ideals. The fact that the number of axionic, non-polynomial scalars in D-dimensional supergravities in Iwasawa gauges is equal to the dimension of the maximal abelian ideal of a solvable Lie
algebra of $E_{11-D (11-D)}$ shows that {\it the existence of the different supergravities in the same dimension D is a generic phenomenon}.

As such, the issue of quantum equivalence between different gauges of a local H-symmetry becomes important: how can we compare the on-shell S-matrix in these different gauges? Especially in the situation that all loop computations are performed using superamplitudes which typically correspond to symmetric gauges of linearized supergravity.

 \section{Noether-Gaillard-Zumino  $Sp(2n_v, \mathbb{R})$ current conservation}\label{Sec:NGZ}

A detailed derivation of the on-shell Noether-Gaillard-Zumino (NGZ) $Sp(2n_v, \mathbb{R})$ current conservation was performed in \cite{Kallosh:2008ic,Kallosh:2011dp} based  on \cite{Gaillard:1981rj,Aschieri:2008ns}. However,  in 
\cite{Kallosh:2011dp}, the focus was also on $\cN=8$ supergravity on-shell superspace  \cite{Brink:1979nt}, 
which allows to build superinvariants, counterterm candidates. on-shell superspace  has maximal duality \E\,, a subgroup of $Sp(56, \mathbb{R})$. Therefore only \E\, current conservation was described in detail. 
In \cite{Hillmann:2009zf,Bossard:2010dq} the focus was also on maximal 4D supergravity and \E\, current conservation.

Here we present the main steps in the derivation of the $Sp(2n_v, \mathbb{R})$ current conservation. We would like to stress the fact that this derivation does not require a choice of a particular form of the $\cN=8$ supergravity or the choice of the gauge. It can be applied therefore to the Lagrangians in \cite{deWit:2002vt,deWit:2002vz,deWit:2005ub,deWit:2007kvg} where the vielbein transforms under $Sp(2n_v, \mathbb{R})$.

At the first step, one derives a so-called {\it NGZ identity} for $Sp(2n_v, \mathbb{R})$, see Sec. 3 in \cite{Kallosh:2011dp}. There is an infinitesimal $Sp(2n,{\mathbb{R}})$ transformation, which acts on $Sp(2n,{\mathbb{R}})$ doublet of vectors field strength 
\be
 \delta_{Sp(2n_v)} \left(\begin{array}{c}F \\G\end{array}\right) = \left(\begin{array}{cc}A & B \\C & D\end{array}\right) \left(\begin{array}{c}F \\G\end{array}\right)\, , \qquad C=C^T, \quad B=B^T , \quad D=-A^T \ ,
\label{GZSp}\ee
where 
\be
 F_{\mu\nu} =\partial_\mu \cA_\nu-\partial_\nu \cA_\mu\, , \qquad \tilde G_{\mu\nu}= 2{\partial L\over \partial F^{\mu\nu}} \ .
\label{NGZ}\ee 
   Duality symmetry on $\varphi^\alpha$ fields is of the form
$
\delta \varphi^\alpha= \Xi^\alpha(\varphi) 
$.
There is a consistency requirement here that the dual field strength  $G$ in the doublet transforms according to the chain rule when expressed as a functional of $F$ and $\varphi$. This consistency condition  is given in the form of NGZ  $Sp(2n,{\mathbb{R}})$ duality identify: it requires that {\it the Lagrangian must transform under duality in a certain way, defined by  NGZ identitiy}
  \cite{Gaillard:1981rj,Aschieri:2008ns,Kallosh:2011dp}
 \be
 {\delta \over \delta F^\Lambda }\Big (  S[F', \varphi']- S[F, \varphi]- {1\over 4} \int (\tilde FCF+ \tilde GBG) \Big)=0. \label{GZ}
 \ee


\noindent  In the $SU(8)$ local version of the theory, the Noether $Sp(2n,{\mathbb{R}})$ current consists of two parts: part from all scalar term ${\cal L}_{\cal V}=-\frac{1}{2}\mbox{Tr}\Big((D_{\mu}{\cal V}){\cal V}^{-1} (D^{\mu}{\cal V}){\cal V}^{-1}\Big) $, this is a standard Noether current,  
 \be 
J^\mu_{\cal V}= {\partial {\cal L}_{\cal V}\over \partial (\partial_\mu {\cal V})}\, \delta \, {\cal V} \ ,
 \ee 
 and part of the scalar-vector terms. This one is not a standard Noether current since  $Sp(2n,{\mathbb{R}})$ duality acts on a vector field strength rather than on a vector. In this case, the corresponding current is a Gaillard-Zumino current \cite{Gaillard:1981rj}
\be
\hat J^\mu_{GZ} \equiv  {1\over 2} \left (\tilde G^{\mu\nu} A \, {\cal A}_\nu -\tilde F^{\mu\nu} C {\cal A}_\nu +\tilde G^{\mu\nu}  B {\cal B}_\nu -\tilde F^{\mu\nu} D {\cal B}_\nu  \right) \ .
\ee
Its divergence cancels the scalar variation of the Lagrangian when equations of motion are satisfied. The  classical Lagrangian provides the conservation of the total current, the Noether current of the scalars, and the Gaillard-Zumino current of vectors:
\be
\partial_\mu J^\mu_{NGZ}= \partial_\mu \hat J^\mu_{GZ}+ \partial_\mu J^\mu_{\cal V}=0 \ .
\ee
The proof that the $Sp(2n,{\mathbb{R}})$ current conservation requires that scalar and vector field equations are satisfied follows from NGZ identity \rf{GZ}.

In \cite{Hillmann:2009zf,Bossard:2010dq} where the Hamiltonian formulation of the maximal 4D supergravity \cite{Cremmer:1979up,deWit:1982bul} was presented, it was pointed out that in the 1st order formalism, a bona fide \E\, Noether current is available. It is likely that this formalism can be applied to  Lagrangians in \cite{deWit:2002vt,deWit:2002vz,deWit:2005ub,deWit:2007kvg} where the vielbein transforms under $Sp(2n_v, \mathbb{R})$. In this way, a bone fide $Sp(2n_v, \mathbb{R})$ current can be derived. Significant progress in this direction was already achieved in \cite{Henneaux:2017kbx}.

\section{GZ duality in even and odd D/2=k : Sp(2n) and SO(n,n)}

The original $Sp(2n_v, \mathbb{R})$ duality  in 4D was discovered by Gaillard and Zumino in \cite{Gaillard:1981rj}. It shows that the scalar-vector part of the Lagrangian of supergravity under infinitesimal $Sp(2n_v, \mathbb{R})$  transformations, where there are $n_v$ vectors, transforms as follows
\be
\delta L= {1\over 4} (FC\tilde F + GB\tilde G)
\label{NGZ1}\ee 
as we see in eq. (2.20) in \cite{Gaillard:1981rj}. Here  the $Sp(2n_v, \mathbb{R})$  transformation acts on vector doublet in a real basis  as shown in eq. 
\rf{GZSp}.
The first term in $\delta L$ in \rf{NGZ1} is a total derivative since $F=d \cA$ and the Bianchi identity $d\tilde F=0$.
But the second term does not vanish off-shell. It does vanish under a condition that
\be
B=0\, \qquad \Rightarrow \qquad \left(\begin{array}{cc}A & 0 \\C & D\end{array}\right)
\ee
This defines a manifest symmetry of the Lagrangian under the so-called electric subgroup $G_e$ of $Sp(2n_v, \mathbb{R})$  with $B=0$.

On-shell equation of motion is  $dG=0$, and we require that $G= d\cB$ and the second term in eq. \rf{NGZ1} also becomes a total derivative. Therefore  $Sp(2n_v, \mathbb{R})$ transformation with $B\neq 0$ relates theories which are not equivalent off shell, but equivalent on-shell.

\noindent Gaillard-Zumino duality symmetry in 4D supergravity \cite{Gaillard:1981rj} was generalized to 6D and 2D in   \cite{Tanii:1984zk}  and  \cite{Cecotti:1988zz}, respectively, and described in general even dimension in  \cite{Andrianopoli:1996ve} and in \cite{Cremmer:1997ct, Tanii:1998px,Samtleben:2008pe,Trigiante:2016mnt}. It is a generalization of the electric-magnetic duality in the Maxwell theory. Is  is available  {\it only in even dimensions} D=2k where there are both electric $F_{\mu_1\dots \mu_k}$ and magnetic $ \tilde F^{\mu_1\dots \mu_k}= {1\over k!} e^{-1} \epsilon^{\mu_1\dots \mu_k \nu_1\dots \nu_k} F_{\nu_1\dots \nu_k}$  $k$-forms. It was observed in \cite{Cecotti:1988zz}  that 
\be
F_1 \tilde F_2 =(-1)^{k}  \tilde F_1 F_2\, , \qquad k={D\over 2} 
\ee
 and therefore a duality operator $X$ acting on a duality doublet 
\be
 \delta_X \left(\begin{array}{c}F \\G\end{array}\right) = X \left(\begin{array}{c}F \\G\end{array}\right)
\label{deltaX}\ee
must satisfy a condition (depending on ${D\over 2} =k$ being even or odd) that 
\be
X^T \Omega= - \Omega X : \qquad \Omega _{k=2p}=\left(\begin{array}{cc}0 & -1 \\1 & 0\end{array}\right)\, ,  \qquad \Omega _{k=2p+1}=\left(\begin{array}{cc}0 & 1 \\1 & 0\end{array}\right) \ .
\label{sym}\ee
Therefore the duality group for $n$ of (D/2-1)-form fields, when ${D\over 2} =k=2p$ is even,  is  $Sp(2n, \mathbb{R})$.  The duality group, for $n$ of (D/2-1)-form fields, when ${D\over 2} =k=2p+1$ being odd, is  $SO(n, n)$. Thus, for maximal supergravities, a duality group, including the GZ duality group, is given in eq. \rf{even}.

It is convenient here to use the ``doubled Lagrangian'' formalism in \cite{Cremmer:1997ct}, which is valid for even dimensions D=2k. In eq. \rf{NGZ1} we have shown the infinitesimal change of $L$ in 4D case. Here we will describe the action of the finite GZ duality on the supergravity Lagrangian in D=2k dimensions.

The form-field-scalar part of the supergravity Lagrangian is
\be
{\cal L}_1= {1\over 2k!} F\cdot \, ^* G + L(\phi)
\ee
where the expression for the k-form $^*G$ can be solved in terms of a k-forms $F$ and $^*F$ using a constraint  imposed on a k-form doublet 
\be
\cH= \left(\begin{array}{c}F \\G\end{array}\right)
\label{cH}\ee
which states that
\be
\cH= \Omega {\cal M} \, ^*\cH 
\label{silver}\ee
It is sometimes called {\it silver rule or twisted self-duality condition} \footnote{It was suggested by H. Nicolai, private communication and \cite{Nicolai:2024hqh},  that 
there is no known way to include the dual magnetic vectors in the
   (on-shell) superspace formulation of N=8 supergravity. Hence there is no known way to deform the twisted self-duality
   constraint in N=8 on-shell superspace.
 If so, there is no counterterm candidate that would be fully
   supersymmetric and invariant under non-linear E7 symmetry. If
   true that would be a strong argument for all order finiteness. 
}

Using the fact that 
\be
^ {**}\cH= (-1)^{k-1} \cH \ ,
\ee
one finds that
\be
^* \cH= \Omega {\cal M} \, ^ {**}\cH = (-1)^{k-1} \Omega {\cal M} \, \cH, \, \quad \cH= (\Omega {\cal M})^2 \, (-1)^{k-1}  \, \cH,  \, \quad (\Omega {\cal M})^2= (-1)^{k-1} I  \ .
\ee
The Lagrangian can also be given in the form
\be
{\cal L}_2=-{1\over 4k!} \cH^T \cM \cH + L(\phi) \ .
\ee
The Bianchi identities  and equations of motion take an elegant form
\be
d\cH=0\, , \qquad  d(\cM ^* \cH)=0
\label{onshellDouble}\ee
and they are invariant under $G_U$-duality symmetry, i. e. \E\, symmetry in 4D and $SO(5,5)$ in 6D and $SL(2, \mathbb{R}) $ in 8D.

Consider now a  GZ duality transformation of the form of a $Sp(2n)$ or $SO(n,n)$ matrix  for even or odd k=D/2,  respectively
\be
\left(\begin{array}{c}F \\G\end{array}\right)'=\left(\begin{array}{cc}U & Z \\W & V\end{array}\right) \left(\begin{array}{c}F \\G\end{array}\right) \ ,
\label{UZWV}\ee
where
\be
U^T V + (-1)^{k-1} W^T Z= V^TU + (-1)^{k-1} Z^T W=I \ ,
\ee
\be
 W^T U+ (-1)^{k-1}U^T W =  V^T Z + (-1)^{k-1}Z^T V =0 \ .
\ee
We can now look at silver rule \rf{silver} transformation under a finite GZ duality, it is duality covariant
\be
\cH'= X \cH \quad \rightarrow \quad  \cH'= (\Omega {\cal M})'  \, ^* X \cH \quad \rightarrow \quad \cH=X^{-1}(\Omega {\cal M})' X * \cH \ ,
\ee
which is satisfied since
\be
X^{-1}(\Omega {\cal M})' X = (\Omega {\cal M}) \quad \rightarrow \quad (\Omega {\cal M})'= X (\Omega {\cal M}) X^{-1} ] .
\ee
Also, the Bianchi identities  and equations of motion after GZ duality transformations are the same
\be
d\cH'=0\, , \qquad  d(\cM ^* \cH)'=0 \ .
\label{oshellGZ}\ee
However, the vector-dependent part of the Lagrangian off-shell changes under GZ transformations as follows
\be
{\cal L}'_1= {\cal L}_1 +  {1\over 2k!} ( F^T U^T W \tilde F + G^T Z^T V \tilde G) \ .
\label{change}\ee
The first term with $F=d\cA$ is a total derivative, so the change of the Lagrangian off shell where $G\neq d\cB$ is up to total derivatives
\be
{\cal L}'_1= {\cal L}_1 +  {1\over 2k!}  G^T Z^T V \tilde G  \ .
\label{offshellGZ}\ee
There is a subgroup of the whole GZ duality group under which the action is invariant. This subgroup is called an electric subgroup $G_e$ and requires that 
\be
 Z^T V =0  \ .
\ee
For infinitesimal transformations, as in 4D example with $k=2$ in \rf{GZSp}, we have $Z\to B; W\to C; U, V\to 1$. Up to a slight change in notation, we see that $Z^T D \to B$ is an infinitesimal change of the action as in \rf{NGZ}. On-shell when 
$G= d\cB$ all inequivalent off shell Lagrangians \rf{offshellGZ} lead to the same equations of motion/Bianchi identities given in \rf{onshellDouble}.

{\it 4D case}

\noindent In maximal 4D supergravity there are four inequivalent  examples of  semisimple electric groups $G_e$, which are manifest symmetries of the Lagrangian   \cite{deWit:2002vt,deWit:2002vz,deWit:2005ub,deWit:2007kvg}
\be
SL(8, \mathbb{R}), \quad E_{6(6)}\times SO(1,1), \quad SL(2, \mathbb{R}\times SO(1,1)\times SL(6, \mathbb{R}), \quad SU^*(8) \ .
\ee
Inequivalent here means that the Lagrangians with these different electric subgroups are related to each other by $Sp(2n_v, \mathbb{R})$ GZ duality transformations, they are not equivalent off-shell. They are only equivalent on-shell. The original actions in \cite{Cremmer:1979up,deWit:1982bul} have $SL(8, \mathbb{R})$ electric group.

In 4D $\cN\geq 5$
 GZ duality group is $Sp(2n_v, \mathbb{R})\supset G_U$ whereas U-duality group $G_U$ is a subgroup of $Sp(2n_v, \mathbb{R})$. In particular for $\cN\geq 5$
\be
\cN=8:  \, \,   G_{GZ}=Sp(56,  \mathbb{R}), \quad  \cN=6:  \, \,  G_{GZ}=Sp(32,  \mathbb{R}), \quad  \cN=5: \, \,  G_{GZ}= Sp(20,  \mathbb{R}) \ .
\ee
For all $\cN\geq 5$ 4D supergravities $G_{GZ} \supset G_U$ i.e.   {\it   GZ duality has many more symmetries than U duality}, see eq. \rf{N568}.

{\it 6D case}

\noindent In 6D maximal supergravity \cite{Tanii:1984zk,Bergshoeff:2007ef} GZ duality symmetry is the same as $G_U$ duality
\be
SO(5,5) \sim E_{5(5)}= G_U \ .
\ee
The electric subgroup, the manifest symmetry of the ungauged supergravity action in  \cite{Tanii:1984zk,Bergshoeff:2007ef} is $GL(5, \mathbb{R})$. There are no other Lagrangians with different electric subgroups related to each other by GZ duality transformations. As we will see later, to have a  Lagrangian different from \cite{Tanii:1984zk,Bergshoeff:2007ef} off shell, one would need a GZ duality with extra symmetries comparatively to $G_U$ duality, but these are absent \footnote{We are grateful to H. Samtleben confirming that in 6D  no different frames are known in ungauged supergravity. The frames are only different in gauged supergravities in 6D, but we are looking here only at ungauged supergravities.} in 6D.

{\it 8D case}

\noindent The coset manifold here is $G/H= {Sl(3, \mathbb{R}\over O(3))}\times {Sl(2, \mathbb{R})\over U(1)}$ so the the U-duality group is $E_{3(3)}= Sl(3, \mathbb{R})\times Sl(2, \mathbb{R})$. The $Sl(2, \mathbb{R}) $ subgroup of the full duality group is the same as the GZ group associated with one  4-form field strength doublet, so that the GZ duality group is $Sp( 2, \mathbb{R})$  \cite{Andrianopoli:1996ve}
\be
G_{GZ} = Sp( 2, \mathbb{R}) \sim Sl(2, \mathbb{R}) \ .
\ee
Here the GZ duality is a symplectic group $Sp( 2n, \mathbb{R})$, however, there is only one field $C_{\mu\nu\rho}$, therefore this symplectic group $Sp( 2, \mathbb{R})$ is small and it is the same as the corresponding subgroup of the U-duality group $Sl(2, \mathbb{R}) $. Thus, for the reason different from the one in 6D,  the maximal duality in 8D is just U-duality, there is no enhancement of duality symmetries, and there are no different symplectic frames and no off-shell different Lagrangians with the same on-shell theory. There is only one ungauged Lagrangian in 8D, the one in  \cite{Salam:1984ft}.


\section{The role of Sp(2n): different Lagrangians, same on-shell S-matrix} 
 
The role of $Sp(56, \mathbb{R})$ group in 4D maximal supergravity with local $SU(8)$ symmetry was revealed in \cite{deWit:2002vt,deWit:2002vz,deWit:2005ub,deWit:2007kvg}. It was explained there that ungauged supergravity Lagrangians are not unique:  the $Sp(56, \mathbb{R})$ rotation generates different Lagrangians, corresponding to different symplectic frames.  Therefore $Sp(56, \mathbb{R})$ rotation is not a symmetry of the Lagrangian. However, there is a symmetry of equations of motion and Bianchi identities, which is a subgroup \E\, of $Sp(56, \mathbb{R})$. {\it Equations of motion and Bianchi identities are the same for all different choices of symplectic frame Lagrangians}. 

Therefore in 4D one can use  $Sp(56, \mathbb{R})$ beyond \E, and the fact that  $E_{7(7)} \subset Sp(56, \mathbb{R})$, to prove the on-shell equivalence between Iwasawa type and symmetric gauges of the local $SU(8)$ symmetry. In 6D where  \EE\, $\sim SO(5,5)$, the symmetric and Iwasawa gauges are not on-shell equivalent, which is consistent with local $SO(5)\times SO(5)$ being anomalous. This supports the existence of 3-loop UV divergence in 6D maximal supergravity \cite{Bern:2008pv}. There is also an anomalous term in the string threshold function associated with 3-loop UV divergence in 6D maximal supergravity \cite{Obers:1999um,Green:2010wi,Pioline:2015yea}. In the context of string threshold functions, the relevant anomalies are known as ``harmonic anomalies''. 
UV divergences in 6D also imply the  local $SO(5)\times SO(5)$ as well as \EE\,  global anomalies in 6D supergravity\cite{Kallosh:2023css,Kallosh:2023dpr,Kallosh:2024lsl}.

After this quick preview, we continue with a more systematic discussion of dualities based on original supergravity papers in 4D and 6D \cite{Cremmer:1979up,Gaillard:1981rj,deWit:1982bul,Tanii:1984zk,Bergshoeff:2007ef}, studies of symplectic frames in 4D \cite{deWit:2002vt,deWit:2002vz,deWit:2005ub,deWit:2007kvg} and reviews of supergravities in various dimensions in  \cite{Andrianopoli:1996ve} and in \cite{Cremmer:1997ct,Tanii:1998px,Samtleben:2008pe,Trigiante:2016mnt}.

Supergravity theories I in dimension D which we will consider contain physical scalar fields that belong to the coset space  $(G/H)_D$.  Here G is a non-compact U-duality group, while H is the maximal compact subgroup of G in dimension D. In the original versions of ungauged supergravities in \cite{Cremmer:1979up,deWit:1982bul,Tanii:1984zk,Bergshoeff:2007ef,Salam:1984ft} the number of scalars is defined by 
the fundamental representation of a group G, and there is also a local H symmetry. This local H symmetry in \cite{Cremmer:1979up,deWit:1982bul,Tanii:1984zk,Bergshoeff:2007ef,Salam:1984ft} has unusual features.
There are no propagating gauge fields, gauge symmetry connections, but the role of the gauge field in local H symmetries is played by the composite scalar dependent connection. Before local H symmetry is gauge-fixed, the global duality symmetry G and the local H symmetry are independent, and fermions transform under H and are neutral in G.

The action with local H symmetry depends on scalars that form an adjoint representation of G, for example, 133 scalars in maximal 4D with $G_4=$\E\,, and 45 scalars in 6D with $G_6$=\EE\,.  
These scalars parametrize a G-valued matrix $\cV(x)$, which transforms by global duality G from the left and by local symmetry H from the right
\be
\cV(x) \to {\bf g}\, \cV(x)\, , \qquad  \cV(x) \to  \cV(x) h^{-1}(x)\, , \qquad {\bf g}\in G\, ,  \qquad h(x) \in H  \ .
\label{old1}\ee
When local H symmetry in supergravities is gauge-fixed, only physical scalars remain; their number is reduced to the number of coordinates in the coset space G/H. Meanwhile, fermions in gauge-fixed theory transform under G symmetry due to a compensating H symmetry transformation, preserving the choice of the gauge.

 In  \cite{deWit:2002vt,deWit:2002vz,deWit:2005ub,deWit:2007kvg} there is an analysis of classical symmetries of the theory with local H symmetry not yet gauge-fixed. We will apply these studies to explicit gauge-fixing local H symmetries of classical theories to understand the relation between different gauges, symmetric and Iwasawa type, in 4D and in 6D.

\subsection{Sp(2n) duality in supergravities with local H symmetry }
 ${Sp(2n_v)} $ Gaillard-Zumino duality \cite{Gaillard:1981rj} acting on vector doublets is shown in eq. \rf{GZSp}. In its original form in  \cite{Gaillard:1981rj} it describes this symmetry for supergravities without local H symmetry. The generalization of ${Sp(2n_v)} $ Gaillard-Zumino duality in the presence of a local H symmetry was realized in \cite{deWit:2002vt,deWit:2002vz,deWit:2005ub,deWit:2007kvg}. 
 
 The standard formalism \cite{Cremmer:1979up,deWit:1982bul,Tanii:1984zk}
starts from a matrix-valued field, ${\cV}(x)$, that belongs to the group G. In 4D it is in the fundamental representation. 
However, in 6D the 16x16 vielbein  ${\cV}_{16}$ of $SO(5,5)$ is an independent object whereas  ${\cV}_{10}$ is a quadratic combination of ${\cV}_{16}$.

The vielbein transforms under rigid G transformations and under local H transformations, as shown in eq. \rf{old1}. G-invariant one-forms are 
\be
{\cV}^{-1} \partial_\mu \cV = {\cal Q}_\mu +{\cal P}_\mu \ .
\label{QP}\ee
H-covariant derivatives are defined as follows $D_\mu \cV= \partial_\mu \cV -\cV {\cal Q}_\mu$ so that the H-covariant scalar dependent tensor tensor ${\cal P}_\mu$ is
\be
{\cV}^{-1} D_\mu \cV = {\cal P}_\mu \ .
\label{P}\ee
The quantities ${\cal Q}_\mu$ and  and ${\cal P}_\mu$  satisfy Cartan-Maurer equations. The scalar Lagrangian is invariant under both rigid G transformations and local H transformations 
\be
{\cal L}_{sc} \propto {1\over 2} \Tr \Big [ D_\mu {\cV}^{-1} \,D^\mu {\cV}\Big ]= {1\over 2} \Tr \Big [ {\cal P}_\mu \, {\cal P}^\mu \Big ] \ .
\label{ac}\ee
We will explain the role of ${Sp(2n_v)} $ Gaillard-Zumino duality in the context of theories with a local H symmetry  here for $\cN=8$ case with $n_v=28$. The   $G_{Sp(56)}$ matrix in a  pseudoreal basis is given by a block matrix with  28x28 elements with 1596 independent parameters
\be
E_{Sp(56)}=\exp \left(
                                        \begin{array}{cc}
                                         \Lambda_{IJ}{}^{KL} & \Sigma _{IJPQ} \\
                                          \Sigma^{MNKL} & \Lambda ^{MN}{}_{PQ} \\
                                        \end{array}
                                      \right)\ .
\ee
where
\be
( \Lambda_{IJ}{}^{KL})^*=  \Lambda^{IJ}{}_{KL}= - \Lambda_{KL}{}^{IJ}\, , \qquad (\Sigma _{IJPQ})^*= \Sigma ^{IJPQ} \ ,
\label{Sp}\ee
and where $I,J = 1,...,8$.
It has a subgroup \E\, with 133 independent parameters. For this subgroup, there are more constraints. In addition to \rf{Sp}, one has to require that
\be
\Lambda_{IJ}{}^{KL}= \delta_{[I}{}^{[K} \Lambda_{J]}{}^{L]}\, , \quad 
\Lambda_I{}^J= - \Lambda ^J{}_I\, , \qquad \Lambda ^I{}_I=0\, , 
\ee 
\be
\quad 
\Sigma_{IJKL}=\pm {1\over 24}  \, \epsilon_{IJKLMNPQ}  \Sigma ^{MNPQ} \ .
\ee
In terms of the real matrix in eq. \rf{GZSp} the restriction on infinitesimal elements of $Sp(56)$ to present a \E\, subgroup are, see for example \cite{Kallosh:2008ic}
\begin{eqnarray}
&& A=\mbox{Re}\Lambda-\mbox{Re}\Sigma\ ,\quad B=\mbox{Im}\Lambda+\mbox{Im}\Sigma\ ,\label{sympA}\\
&& C=-\mbox{Im}\Lambda+\mbox{Im}\Sigma\ ,\quad D=\mbox{Re}\Lambda+\mbox{Re}\Sigma\ .\label{sympC}
\end{eqnarray}
Thus the manifest symmetry of the Lagrangian under an electric subgroup of $Sp(56)$ means that 
\be
B=0 \qquad \Rightarrow \qquad \mbox{Im}\Lambda+\mbox{Im}\Sigma=0 \ .
\ee
 The classical action \cite{deWit:1982bul} with a local SU(8) symmetry before this symmetry is gauge fixed depends on 133 scalars represented by a  56-bein.  It is an element of \E\,  
\begin{eqnarray}\label{V}
{\cal V}(x) =\left(
                                        \begin{array}{cc}
                                          u^{ij} {} _{IJ}(x)&\, \, \,  -v_{klIJ}(x) \\
                                          \\
                                          - v^{ijKL} (x)&\, \, \,   u_{kl}{}^{KL} (x)\\
                                        \end{array}
                                      \right)\,                          
                                      \end{eqnarray}  
The indices $I,J$ and $i,j$ take the values 1,...,8, so that there are 28 anti-symmetrized index pairs representing the matrix indices of $\cV $.                                    
                                      
In supergravities with local H symmetry, one would expect different  Lagrangians,
which are manifestly invariant under different subgroups of \E\,. To study these different Lagrangians a constant $Sp(56, \mathbb{R})$ matrix E is introduced in \cite{deWit:2002vt,deWit:2002vz,deWit:2005ub,deWit:2007kvg} where $A,B = 1,...,8$
\be
E=\left(\begin{array}{cc}U_{IJ}{}^{AB} & \, \, \, V_{IJCD} \\
\\
V^{KLAB} &\, \,  \, U^{KL}{}_{CD}\end{array}\right)
\label{E}\ee
Matrix $E$ belongs to a double quotient $E_{7(7)}\backslash Sp(56;\mathbb{R})/GL(28)$, which defines the embedding of the 28 vector fields into the 56-bein, as discovered in \cite{deWit:2002vz}.

This constant matrix $E$ was called $ {\cV}^{0}$ in \cite{deWit:2007kvg} to stress that the $x$ dependent vielbein in eq. \rf{V} is defined up to a choice of its constant value. Depending on the choice of this constant part $ {\cV}^{0}$, the action takes different forms specified by the choice of the  constant matrix $E_{7(7)}\backslash Sp(56;\mathbb{R})/GL(28)$.

One defines a new 56-bein, which is  no longer a group element of \E    
 \be
 \widehat {\cal V} (x)= E^{-1} {\cal V}(x)=\left(
                                        \begin{array}{cc}
                                          u^{ij} {} _{AB}(x)&\, \, \,  -v_{klCD}(x) \\
                                          \\
                                          - v^{ijAB} (x)&\, \, \,   u_{kl}{}^{CD} (x)\\
                                        \end{array}
                                      \right)\, .
\label{hat} \ee    
The local SU(8) covariant  and $G_{Sp(56), \mathbb{R}}$ invariant doublet  is 
 \be
\left(\begin{array}{c}\cF^+_{\mu\nu \, ij} \\
\\
\cO^{+kl}_{\mu\nu}\end{array}\right) = \cV^{-1} E \left(\begin{array}{c}F^+_{1 \mu\nu \, AB} \\
\\
F^{+AB}_{2 \mu\nu}\end{array}\right) \ .
\label{new} \ee 
It presents a graviphoton   $\cF^+_{\mu\nu \, ij}$ and a bilinear of fermions  $\cO^{+kl}_{\mu\nu}$.                               
 Here  $F_{1}$     and $F_2$ are certain $\pm$ combination of a 2-form $F$ in the action and its dual  $G$, as in eq. \rf{NGZ} and ${\cal O}$ is a function of fermions.        
 
 The definition of a local SU(8) doublet, \E\, invariant   was given in \cite{Cremmer:1979up,deWit:1982bul}.     For example, in the notation of   \cite{deWit:1982bul}, the local  SU(8) doublet, \E\, invariant, is   given in eq. (2.23) there in the form  \be
 \left(\begin{array}{c}\cF^+_{\mu\nu \, ij} \\
\\
\cO^{+kl}_{\mu\nu}\end{array}\right)=
\cV ^{-1} \left(\begin{array}{c}F^+_{1 \mu\nu IJ} \\
\\
F^{+}_{2 \mu\nu}{}^{IJ}\end{array}\right) \ . \label{old}\ee    
The way to see that the LHS is a local SU(8) doublet  and \E\, invariant is to remember that 
\be
\cV ^{-1} \to h(x) \cV ^{-1} (g_{_{E_{7(7)}}})^{-1}\, ,  \left(\begin{array}{c}F^+_{1 \mu\nu } \\
\\
F^{+}_{2 \mu\nu}\end{array}\right)\to g_{_{E_{7(7)}}}
 \left(\begin{array}{c}F^+_{1 \mu\nu } \\
\\
F^{+}_{2 \mu\nu}\end{array}\right)\, \,   {\rm and} \,  \, \left(\begin{array}{c}\cF^+_{\mu\nu \, ij} \\
\\
\cO^{+kl}_{\mu\nu}\end{array}\right)\to h(x)\left(\begin{array}{c}\cF^+_{\mu\nu \, ij} \\
\\
\cO^{+kl}_{\mu\nu}\end{array}\right)
\label{expl} \ee    
The new story with $Sp(56, \mathbb{R})$ symmetry in \rf{new} can be explained by analogy with \rf{expl} as follows
\bea
&&\cV ^{-1} \to h(x) \cV ^{-1} (g_{_{E_{7(7)}}})^{-1}\, ,\quad  E\to g_{_{E_{7(7)}}} E \, (g_{_{Sp(56)}})^{-1}\, ,  
\eea
\bea
&& \left(\begin{array}{c}F^+_{1 \mu\nu } \\
 \\
F^{+}_{2 \mu\nu}\end{array}\right)\to g_{_{Sp(56)}} \left(\begin{array}{c}F^+_{1 \mu\nu } \\
\\
F^{+}_{2 \mu\nu}\end{array}\right)\, \,  \quad  {\rm and} \quad \,  \, \left(\begin{array}{c}\cF^+_{\mu\nu \, ij} \\
\\
\cO^{+kl}_{\mu\nu}\end{array}\right)\to h(x)\left(\begin{array}{c}\cF^+_{\mu\nu \, ij} \\
\\
\cO^{+kl}_{\mu\nu}\end{array}\right) \ .
\label{expl1} \eea  
Thus, the matrix E in \rf{E} which was absent in early versions of supergravities \cite{Cremmer:1979up,deWit:1982bul} and was introduced in \cite{deWit:2002vt,deWit:2002vz,deWit:2005ub,deWit:2007kvg}   interpolates between \E\, and  $Sp(56, \mathbb{R})$ and in this way allows the vector doublet to transform under  $Sp(56, \mathbb{R})$, as prescribed by Gaillard Zumino duality.   It may also be useful to note that
\be
\cV^{-1} E = (\hat \cV)^{-1} \to h(x) (\hat \cV)^{-1} (g_{_{Sp(56)}})^{-1} \ .
\ee                        
 Together with \rf{expl1}
it clearly shows that in the LHS of eq. \rf{new} the local SU(8) doublet is $Sp(56, \mathbb{R})$ invariant.

Note that the matrix E, being a constant matrix,  drops from $\cal Q$ and $\cal P$
\be
{\hat\cV}^{-1}  \partial_\mu \hat \cV = {\cV}^{-1} \partial_\mu \cV=
{\cal Q}_\mu +{\cal P}_\mu
\label{QP1}\ee
Therefore the scalar Lagrangian is invariant under both local H transformations and global $Sp(56, \mathbb{R})$. Using notations of \cite{deWit:2007kvg},
\be\label{sc}
  e^{-1} {\cal L}_{sc}  =- {1\over 12} |cP_\mu ^{ijkl}|^2 \ .
\ee
The vector/scalar action is
\begin{eqnarray}
  \label{newvs}
  e^{-1} \mathcal{L}_{\rm vector} =
-{i\over 4} [ \cN_{\Lambda\Sigma} 
\, ({F}_{\mu\nu}^+{}^{\Lambda})\, ({F}^{+\mu\nu\Sigma})  -
\overline \cN_{\Lambda\Sigma} \, ({F}_{\mu\nu}^-{}^{\Lambda})\, 
 ( {F}^{-\mu\nu\Sigma})]
\end{eqnarray}
where
\be
(u^{ij} {} _{AB}+ v^{ijAB}) {\cal N}_{AB, CD} = (u^{ij} {} _{CD}- v^{ijCD}) \ .
\ee
Under $Sp(56, \mathbb{R})$ transformation, in general, the Lagrangian will change into a Lagrangian in a different symplectic frame. Only under the electric subgroup of  $Sp(56, \mathbb{R})$ the Lagrangian will be invariant.

First, consider the action of the $Sp(2n_v,\mathbb{R})$ symmetry on the off-shell Lagrangian in \cite{deWit:2007kvg}.
The vector part is given in eq. \rf{UZWV} and duality transformation is defined by a real constant matrix $\left(\begin{array}{cc}U & Z \\W & V\end{array}\right)$ which leaves the skew-symmetric matrix
$\Omega_{MN}$ invariant and the scalar dependent vector metric transforms as follows
\bea
\mathcal{N}_{\Lambda\Sigma}'=
  (V \mathcal{N} + W )_{\Lambda\Gamma} \,
  [(U + Z \mathcal{N})^{-1}]^\Gamma{}_\Sigma  \ .
\eea
The manifest duality symmetry of the action in a given symplectic frame is an electric subgroup of $Sp(2n_v,\mathbb{R})$ of the form
\be
G_e= \left(\begin{array}{cc}U & 0 \\W & V\end{array}\right) \ .
\label{el}\ee
This enforces a new choice of a coset representative, like $\phi'$ instead of $\phi$.

The new Lagrangian after the $Sp(2n_v,\mathbb{R})$ transformation is
\begin{eqnarray}
  \label{new1}
  e^{-1} \mathcal{L}'_{\rm vector} =
-{i\over 4} [ \cN'_{\Lambda\Sigma} 
\, ({F}_{\mu\nu}^+{}^{\Lambda})'\, ({F}^{+\mu\nu\Sigma})'  -
\overline \cN'_{\Lambda\Sigma} \, ({F}_{\mu\nu}^-{}^{\Lambda})'\, 
 ( {F}^{-\mu\nu\Sigma})'] \ .
\end{eqnarray}
As we already explained around eq.  \rf{change}, the difference between these Lagrangians is
\be
{\cal L}'- {\cal L} \propto   ( F^T U^T W \tilde F +
G^T Z^T V \tilde G) \ .
\label{change1}\ee
The first term with $F=d\cA$ is a total derivative, so the change of the Lagrangian off shell where $G\neq d\cB$ is up to total derivatives
\be
{\cal L}'- {\cal L} \propto   
G^T Z^T V \tilde G \qquad {\rm off} \, {\rm shell}  \ .
\label{offshellGZ1}\ee
However, on-shell when $G= d\cB$
\be
{\cal L}'= {\cal L}  \qquad  {\rm on} \, {\rm shell}  \ .
\label{onshellGZ1}\ee
Thus, off-shell, the action really changes after an $Sp(2n_v,\mathbb{R})$ transformation. In particular, it brings the action to a new symplectic frame. The change in cases that the $Sp(2n_v,\mathbb{R})$ transformation does not belong to the double quoitient $G_U (\mathbb{R})  \backslash Sp(2n_v; \mathbb{R})/GL(2n_v)$ can be eliminated by a change of scalar and vectors variables. But in case of $G_U (\mathbb{R})  \backslash Sp(2n_v; \mathbb{R})/GL(2n_v)$, the change in the action cannot be eliminated.

Thus,  $Sp(2n_v,\mathbb{R})$ duality transformations define the equivalence classes of Lagrangians that lead to the same field equations and Bianchi identities. This construction indicates that different choices of gauges in various symplectic frames might lead to equivalent physical observables in perturbative supergravity.

\subsection{Double quotients in 4D, 6D, 8D}\label{Sec:double}

 An important development of $Sp(2n_v, \mathbb{R})$ symmetry was in 4D ungauged supergravity Lagrangians with G/H coset space and with local H symmetry in  \cite{deWit:2002vt,deWit:2002vz,deWit:2005ub,deWit:2007kvg}. Their result was
that for ungauged 4D supergravity, the set of Lagrangians that cannot be mapped to each other by local field redefinitions is identified with the double quotient space
\be
E^{{4D}}=G_U (\mathbb{R})  \backslash Sp(2n_v; \mathbb{R})/GL(2n_v) \ .
\label{DQ}\ee
For example, the relevant quotient in maximal 4D supergravity is
\be
E_{_{\cN=8}}^{{4D}}=E_{7(7)}(\mathbb{R}) \backslash Sp(56, \mathbb{R}) / GL(28, \mathbb{R}) \ .
\label{N8}\ee
Here the left quotient, a continuous  $E_{7(7)}(\mathbb{R})$, corresponds to a local redefinition of the scalar fields, it is not the  \E\,  duality, which also acts on vector fields. The right quotient $GL(28, \mathbb{R})$ corresponds to a local redefinition of the 28 vector fields. The resulting theories, codified by the matrix $E$ which belongs to $E_{7(7)} \backslash Sp(56; \mathbb{R})/GL(28)$ are equivalent at the level of the equations of motion!

The clarification of the status of $Sp(2n_v, \mathbb{R})$ relevant to $\cN\geq 5$ 4D supergravities is the following. These theories have non-unique Lagrangians (not related by local field redefinition) codified by the matrix $E$ which belongs to $G (\mathbb{R})\backslash Sp(2n_v,\mathbb{R})/GL(n_v , \mathbb{R})$. Here the relevant dualities are $G= SU(1,5), SO^*(12), E_{7(7)}$ for $\cN=5,6,8$, respectively.
But all these inequivalent Lagrangians for each $\cN$ are equivalent at the level of the equations of motion!

 These results are well known in the context of gauged supergravities and embedding tensor formalism, see for example \cite{DallAgata:2014tph}. They are 
also supported in 
 Hamiltonian formalism analysis in \cite{Henneaux:2017kbx}. There is an agreement on the fact that there is an equivalence between classical theories satisfying equations of motion when symplectic transformations with matrices which belong to $G (\mathbb{R})\backslash Sp(2n_v,\mathbb{R})/GL(n_v , \mathbb{R})$ are performed. 
 
 The counting of elements of a quotient matrix goes as follows in the maximal case: dim $Sp(2n) $ is $2n^2+n$, which in case of $n=28$ is 1596 and 
 dim $GL(28)$ is 784. So we have for the quotient $E_{7(7)}(\mathbb{R}) \backslash Sp(56, \mathbb{R}) / GL(28, \mathbb{R})$
 \be
{\rm dim} [E_{_{\cN=8}}^{4D}]= {\rm dim} \, \, [E_{7(7)}(\mathbb{R}) \backslash Sp(56, \mathbb{R}) / GL(28, \mathbb{R})] : \qquad 1596-133-784=679 
 \ee
many independent parameters. In cases of $\cN=5,6$ analogous counting shows that the quotients $G (\mathbb{R})\backslash Sp(2n_v,\mathbb{R})/GL(n_v , \mathbb{R})$ are highly nontrivial. For example in $\cN=5$ case
\be
E_{_{\cN=5}}^{{4D}}=SU(1,5)(\mathbb{R}) \backslash Sp(20, \mathbb{R}) / GL(10, \mathbb{R}) : \qquad 210-35-100=75
\label{N5}\ee
In 6D, the action of maximal supergravity is given in \cite{Tanii:1984zk,Bergshoeff:2007ef}. Maximal Gaillard-Zumino duality is $SO(5,5)\sim$ \EE\,. The analog of the quotient in eq. \rf{DQ}  in 6D would be the following quotient
\be
E_{_{\cN=8}}^{6D}=E_{5(5)}(\mathbb{R}) \backslash SO(5,5, \mathbb{R}) / G_{v,t}( \mathbb{R}) \ .
\label{N81}\ee
We take out from maximal Gaillard-Zumino duality $SO(5,5)$ a possibility to redefine the 45 scalars in the 16-bein in \cite{Tanii:1984zk} which is in the fundamental of $SO(5,5)$. This is the meaning of the left quotient $E_{5(5)}$. And we can redefine the 16 vector fields and 5 tensor fields in the action, which is the meaning of the right quotient $G_{v,t}( \mathbb{R})$. However, it is clear already at the stage of the left quotient that nothing is left in $E_{5(5)}(\mathbb{R}) \backslash SO(5,5, \mathbb{R})$, since both of these have 45 parameters
and therefore
\be
E_{_{\cN=8}}^{6D}={\rm I}
\label{n8D6}\ee
This means that one can use an $SO(5,5)$ transformation to produce a Lagrangian different from the one in \cite{Tanii:1984zk}. But as long as a local $SO(5)\times SO(5)$ symmetry is present and the Lagrangian depends on 45 scalars in the vielbein, the resulting Lagrangian can be brought back to the form it has in  \cite{Tanii:1984zk}: the left quotient \EE\,   is taking away the effect\footnote{I am grateful to H. Samtleben for confirming that the relevant quotient in 6D ungauged supergravity is trivial, so there are no different frames. The remark in \cite{Bergshoeff:2007ef} that ``there is always a frame, which may be reached by an $O(5, 5)$ rotation from Tanii's Lagrangian'' was meant with regard to the gauged theory described by different Lagrangians (related by SO(5,5) rotation).  } of a Gaillard-Zumino maximal duality symmetry $SO(5,5)$ in 6D.

\section{ Maximal supergravity 4D symplectic and 6D orthogonal  duality frames}
The role of $Sp(56; R)$ symmetry in classical maximal supergravity is to relate Lagrangians with local $SU(8)$ symmetry but with inequivalent manifest symmetry (electric) groups.  These ones might be different from the original Lagrangians in \cite{Cremmer:1979up,deWit:1982bul}.  The difference between them involves the choice of the matrix E in eq. \rf{hat}. 

There are four inequivalent cases studied in  \cite{deWit:2002vt} in detail. The name of the symplectic frame is identified with the name of the manifest electric symmetry of a given Lagrangian. We will be only interested in two of these four frames, the original one with manifest $SL(8, \mathbb{R})$ symmetry and the one with manifest $E_{6(6)}\times SO(1,1)$ symmetry related to 5D maximal supergravity and 4d supergravity II.

\begin{enumerate}
  \item {\bf ${\bf SL(8, \mathbb{R})}$ frame}
  
The standard (ungauged) Lagrangian in 4D \cite{Cremmer:1979up, deWit:1982bul} has manifest  $G_e=SL(8, \mathbb{R})$ symmetry which is a subgroup of \E\,.  The duality representation ${\bf 56}$ of ${\rm E}_{7(7)}$ and its adjoint representation ${\bf 133}$ branch with respect to ${\rm SL}(8,\mathbb{R})$ as follows:
    \begin{align}
    {\bf 56}&\rightarrow {\bf 28}+{\bf 28}'\,,\nonumber \\
    {\bf 133}&\rightarrow {\bf 63}+{\bf 70}\,.\label{sl8branch}
    \end{align} 
 The matrix E in eq. \rf{hat} is trivial 
 \be
 E= E_{SL(8, \mathbb{R})}={\rm I}
 \ee
 The maximal symmetry of the off-shell Lagrangian is $SL(8, \mathbb{R})$  in this frame, but the combined field equations and Bianchi identities have \E\,  as a symmetry group. 
 
  \item ${\bf E_{6(6)}}$ {\bf frame}
  
The electric subgroup of \E\, which is a manifest symmetry of the Lagrangian in this frame, is  $G_e=E_{6(6)}\times SO(1,1)$ and  ${\mathfrak e}_{7(7)}$ algebra has a grading under ${\mathfrak so}(1,1)$ 
\be
{\mathfrak e}_{7(7)} = {\mathfrak l}_0 +{\mathfrak l}_{+2} + {\mathfrak l}_{+2}
\ee

All generators of this $G_e$ are presented in \cite{Andrianopoli:2002mf} and in \cite{deWit:2002vt}. With respect to ${\rm E}_{6(6)}\times {\rm SO}(1,1)$
the ${\bf 56}$ and the adjoint of ${\rm E}_{7(7)}$ decompose  as follows:
\begin{eqnarray}
{\bf 56}&\rightarrow&   {\bf
1}_{-3}+{\bf 27}'_{-1}+{\bf 1}_{+3} +    {\bf 27}_{+1} \,,
\label{56e6}\\
{\bf 133}&\rightarrow&{\bf 27}_{-2}+ {\bf 1}_0+{\bf 78}_0+{\bf 27}'_{+2}\,.\label{133e6}
\end{eqnarray}
Here {\bf 28} vectors in the action split into ${\bf
1}_{-3}$ from a 5D metric and  27 vector fields of 5D in ${\bf 27}'_{-1}$ of the electric group. The ${\bf 78}_0$ scalars are related to generators of $E_{6(6)}$ out of \E\,. The ${\bf 27}'_{+2}$ scalars are  27 axionic scalars originating from the five-dimensional vector fields through the Kaluza-Klein reduction. The  ${\bf 1}_0$ is related to a radius of the 5th dimension. 

In  \cite{Andrianopoli:2002mf}  there is  an explicit form of the action of ${\mathfrak l}_{\pm 2}$ on 
electric/magnetic field strengths
\be
\delta \left(\begin{array}{c}F^\Lambda \\F \\G_\Lambda \\G\end{array}\right) = \left(\begin{array}{cccc}0^\Lambda{}_\Sigma \, \, & -t^{'\Lambda} \, \, & d^{\Lambda \Sigma \Gamma} t_\Gamma \, \, & 0^\Lambda \\-t_\Sigma \, \, & 0 \, \, & 0_\Sigma \, \, & 0 \\d_{\Lambda \Sigma \Gamma} t^{'\Gamma}\, \,  & 0_\Lambda \,\,  & 0_\Lambda{}^\Sigma \, \,  & t_\Lambda  \\0_\Sigma \, \,  & 0 \, \, & t^{' \Sigma} \, \,  & 0\end{array}\right)\, \left(\begin{array}{c}F^\Lambda \\F \\G_\Lambda \\G\end{array}\right)
\ee
Here $\Lambda=1,\dots, 27$, $t_\Lambda$ and $t^{'\Lambda}$ are the parameters of the transformation,
$d_{\Lambda \Sigma \Gamma}$   is the symmetric invariant tensor of the representation ${\bf 27}_{+1}$ of $E_{6(6)}$  and $d^{\Lambda \Sigma \Gamma}$  of ${\bf 27}'_{-1}$. These define 
 the cubic invariant
 of $E_{6(6)}$.

 The matrix E in eq. \rf{hat} is a nontrivial  element of the quotient 
  \be
 E= E_{E_{6(6)}} \Rightarrow  E_{7(7)}(\mathbb{R}) \backslash Sp(56, \mathbb{R}) / GL(28, \mathbb{R})
\label{quo}\ee
It relates the theory in the $ E_{6(6)}$ frame to the original  one in \cite{Cremmer:1979up, deWit:1982bul} in $SL(8, \mathbb{R})$ frame. A detailed explanation of the $E_{E_{6(6)}} $ matrix is given in  \cite{deWit:2002vt}.

\end{enumerate}
Two more symplectic frames in maximal 4D supergravity were presented in \cite{deWit:2002vt}, one is $SL(2, \mathbb{R})\times SO(1,1) \times SL(6, \mathbb{R})$ frame, and the other is $SU^*(8)$. The $SU^*(8)$ basis and related gaugings of maximal 4D supergravity were discovered in \cite{Hull:2002cv}.  In \cite{deWit:2002vt}, these were also viewed as a consequence of the double quotient $E_{7(7)}(\mathbb{R}) \backslash Sp(56, \mathbb{R}) / GL(28, \mathbb{R})$ transformation starting from the $SL(8, \mathbb{R})$ basis.

In 6D the maximal Gaillard-Zumino duality group is $SO(5,5)$, therefore there is no known analog of different 6D Lagrangians related by specific \EE\,  transformations. The $SO(5,5)$ transformation can be absorbed into the redefinition of the vielbein. Therefore there is no known  relation between 6D action in \cite{Tanii:1984zk} and in 7D$\to$ 6D action  in \cite{Cowdall:1998rs}.


 \section{ 4D classical $Sp(2n_v, \mathbb{R})$ on-shell symmetry }
 
 In maximal 4D supergravity we can start with the classical action in $SL(8, \mathbb{R})$ basis in 
\cite{Cremmer:1979up, deWit:1982bul} and proceed with the gauge-fixing following \cite{Cremmer:1979up, deWit:1982bul,Kallosh:2008ic} and  \cite{Kallosh:2024rdr}.  The classical 56-bein in eq. \rf{V} is an \E\, element which depends on 133  scalars.

We can gauge-fix it in {\it a symmetric gauge} where all  70 physical scalars enter the action non-polynomially  \be
{\cal V}^{sym}_{g.f} \qquad \to  \qquad {\cal V}_{gf} = {\cal V}_{gf} ^{\dagger} = e^{X}= \begin{pmatrix}\cosh \phi \bar \phi& \phi \frac{\sinh \bar \phi \phi}{\bar \phi \phi}\cr\bar  \phi \frac{\sinh \phi \bar\phi}{ \phi \bar\phi}&\cosh \bar \phi \phi\end{pmatrix}\,,  \qquad  X= \left(\begin{array}{cc}0 & \phi_{ijkl} \\\bar \phi^{mnpq} & 0\end{array}\right)  \,.
\ee
Here the self-dual scalars $\phi_{ijkl}= \pm {1\over 4!} \epsilon_{ijklpqmn}\bar \phi^{pqmn}$ transform in the 35-dimensional representation of $SU(8)$. The gauge-fixed theory has manifest global $SU(8)$ symmetry.

The Iwasawa and partial Iwasawa gauge gauges in 4D maximal supergravity were described in  \cite{Kallosh:2024rdr}, based on type II supergravity set of scalars in \cite{Andrianopoli:2002mf}. Now we can describe it as follows: one starts with the classical action in \cite{deWit:2002vt} with local   $SU(8)$ symmetry which is made after first performing an $Sp(56, \mathbb{R})$ GZ duality transformation. This brings the classical action to the ${\bf E_{6(6)}}$ basis. The corresponding classical action is given in  \cite{deWit:2002vt}, where it is characterized by a 56-bein
$
 \widehat {\cal V} (x)= E^{-1} {\cal V}(x) \ ,
$   
where $E$ in general is defined in eq. \rf{E}. In this basis 
the  group $G_e$ contains the semisimple group $E_{6(6)} \times SO(1,1) \subset E_{7(7)}$ as a subgroup. The features of this basis are described in detail in  \cite{deWit:2002vt}, which leads to the statement that the matrix $E$ is a nontrivial element of the double quotient $E_{7(7)}(\mathbb{R}) \backslash Sp(56, \mathbb{R}) / GL(28, \mathbb{R})$.

Specifically, it is explained in  \cite{deWit:2002vt} that the ${\bf E_{6(6)}}$-basis involves a matrix $E $ such that ``the Lagrangian coincides with the Lagrangian that one obtains upon reduction to four dimensions of five-dimensional ungauged maximally supersymmetric supergravity''.
 
We need to clarify here the meaning of this statement. The Lagrangian of maximal 4D supergravity in the form with a local $SU(8)$ symmetry is first taken into an   ${\bf E_{6(6)}}$-basis  \cite{deWit:2002vt} where the 56-bein is given by  $\widehat {\cal V} (x)$. This is to be compared with the 5D supergravity reduced to 4D.

In  \cite{Sezgin:1981ac}, a dimensional reduction of the massless maximal supergravity in  5D  to gauged supergravity in 4D was performed. In  \cite{Andrianopoli:2002mf}, the limit to vanishing gauge couplings in \cite{Sezgin:1981ac} in 4D was studied, and in this way, the Lagrangian of ungauged maximal supergravity was obtained. Since in \cite{Sezgin:1981ac}  the local $USp(8)$ symmetry was gauge-fixed already in 5D before dimensional reduction, the 4D maximal ungauged supergravity action with manifest $G_e$ being $E_{6(6)} \times SO(1,1) \subset E_{7(7)}$ does not have local internal bosonic symmetries, they are all gauge fixed.
  
  To compare it with the Lagrangian in  \cite{deWit:2002vt} where the 56-bein is given by  $\widehat {\cal V} (x)$ where there is still a local $SU(8)$ symmetry, we have to perform gauge-fixing of this local $SU(8)$ symmetry. A natural gauge-fixing condition here is the partial Iwasawa gauge so that the theory in \cite{deWit:2002vt} coincides with the theory in  \cite{Andrianopoli:2002mf}. 
  
\be
\widehat {\cV}_{gf}= e^{a^\l t_\l} \, e^{\phi^{abcd} K_{abcd}} \, e^{\sigma D} \ .
\ee
The expressions for the 56x56 matrices, operators $D, t_\l$ are given in eq. (4.26) in \cite{Trigiante:2016mnt} and 27 $t_\l$ form an algebra $[t_\l, t_\delta] =0$. But the  coset representative for the 42 scalars $\phi^{abcd} $ in ${E_{6(6)}\over USp(8)}$ coset space is not taken in  the solvable parametrization. Instead, it is taken in a symmetric gauge in the form   $e^{\phi^{abcd} K_{abcd}}$.  It is the one which was chosen in \cite{Andrianopoli:2002mf},  it corresponds to a symmetric gauge in 5D in  \cite{Sezgin:1981ac} and leads to a partial Iwasawa gauge in 4D in \cite{Andrianopoli:2002mf}.

The central part of the argument about the gauge equivalence comes from the fact that maximal 4D supergravity in ${\bf SL(8, \mathbb{R})}$ frame (before gauge-fixing local $SU(8)$ symmetry) and maximal 4D supergravity in ${\bf E_{6(6)}}$ frame are related by $Sp(56, \mathbb{R})$ symmetry so that the Lagrangians are different off shell but lead to the same equations of motion.

Therefore the classical statement that change of the symplectic frame does not affect classical equations of motion can be interpreted as the statement that the S-matrix does not depend on the choice of the gauge of the local H-symmetry: symmetric gauge, natural in ${ SL(8, \mathbb{R})}$ frame and partial Iwasawa gauge, natural in ${ E_{6(6)}}$  frame. 

In short, classically, it was established in  \cite{deWit:2002vt,deWit:2007kvg} that supergravity I is gauge-independent on-shell.  Moreover, the gauge-fixed supergravity II in \cite{Andrianopoli:2002mf} is equivalent on-shell to supergravity I in the ${ E_{6(6)}}$  frame in the partial Iwasawa gauge.

To promote this classical argument to a quantum level, one has to involve a path integral and $Sp(2n_v, \mathbb{R})$  NGZ current conservation. However, in Lorentz covariant supergravities 
it is not obvious how to proceed since in the path integral we integrate over the vector fields $A_\mu$ but $Sp(2n_v, \mathbb{R})$ duality symmetry acts on a field strength $F_{\mu\nu}=\partial_{[\mu} A_{\nu]}$,  and therefore dualities are not well defined as a change of the variables in the Lorentz covariant path integral. 

To be able to treat duality symmetry transformation as a change of variables in the path integral, we switch to 1st-order formalism where $Sp(2n_v, \mathbb{R})$ duality symmetry of the S-matrix can be viewed as a canonical change of variables in the Hamiltonian path integral.   

\section{Duality symmetry and path integral }\label{Sec:8}

A possibility to promote the construction in  \cite{deWit:2002vt,deWit:2007kvg} to a quantum path integral would be to use the first-order Hamiltonian formulation of dualities developed in \cite{Henneaux:2017kbx} and earlier work in \cite{Hillmann:2009zf,Bossard:2010dq,Bunster:2010kwr}. In general,  the Hamiltonian path integral  was developed by Faddeev \cite{Faddeev:1969su} and by Fradkin-Tyutin-Batalin-Vilkovisky
\cite{Fradkin:1970pn,Fradkin:1973wke,Batalin:1977pb,Fradkin:1977hw}. A comprehensive study of the path integral with gauge degrees of freedom in the Hamiltonian form is presented in  \cite{Henneaux:1985kr}.

In the Hamiltonian formulation, duality transformations are local and act on  3-dimensional  vector field doublets.
We will restrict ourselves to the bosonic part of 4D maximal supergravity, and moreover, we will neglect interaction with gravity, as in \cite{Henneaux:2017kbx}. This means that we will study only part of the 4D maximal supergravity action depending on vector and scalar fields.  These are the only fields that are not inert under duality transformations, whereas the gravitational field is inert.

The scalar part of the action
 is 
\be
  e^{-1}{\cal L}_{scalar}  =-{1\over 12} | {\cal P}_\mu^{ijkl}|^2\, ,  \qquad {\hat \cV}^{-1} \partial_\mu \hat{\cV} - {\cal Q}_\mu = {\cal P}_\mu
\label{scalar}\ee
 where the vielbein ${\hat \cV}$ \cite{deWit:2002vt} can describe any symplectic frame as well as any choice of the coset representative so that the local H-symmetry is gauge-fixed. This action is invariant under $Sp(2n_v, \mathbb{R})$ duality symmetry acting on scalars. There is no need here to transform to a Hamiltonian formalism. This we have to do only in the part where vectors interact with scalars.

Consider the vector scalar action in \cite{Trigiante:2016mnt} in the form where a choice of a symplectic frame was made, as well as a gauge-fixing, a choice of the coset representative, was made. We take it in the form
\be
  e^{-1}\cL_{vector} = - \frac{1}{4} \mathcal{I}_{IJ}(\phi) F^I_{\mu\nu} F^{J\mu\nu} + \frac{1}{8} \mathcal{R}_{IJ}(\phi)\, \varepsilon^{\mu\nu\rho\sigma} F^I_{\mu\nu} F^J_{\rho\sigma} 
\label{vec}\ee
which makes it easier to use the transition to the 1st order formalism, following  \cite{Henneaux:2017kbx}. Here  $ F^I_{\mu\nu} = \partial_\mu A^I_\nu -  \partial_\nu A^I_\mu $.

From this Lagrangian one can derive the Hamiltonian and define the generalized action from which equations of motion can be derived of the type discussed in \cite{Faddeev:1969su,Fradkin:1970pn, Fradkin:1973wke,Batalin:1977pb,Fradkin:1977hw,Henneaux:1985kr,Grigorian:1992sx}
\be
S(q,p, \lambda) = \int ( p_i \dot q^i - H(q,p) -\lambda_a \phi^a(q,p)) dt
\label{Fad}\ee 
The set of constraints can be realized as a set of equations of motion over the Lagrange multipliers $\lambda_a$. Note that we do not have local symmetries in our Lagrangian as the local H-symmetry is gauge-fixed in some unitary gauge by the choice of the coset representative. There is only an Abelian gauge symmetry acting on the vector fields. Therefore there is no need for additional constraints $\chi^a(p,q)$, like in  cases in \cite{Faddeev:1969su,Fradkin:1970pn,Fradkin:1973wke,Batalin:1977pb,Fradkin:1977hw,Henneaux:1985kr,Grigorian:1992sx} suitable for non-Abelian and gravitational  Lagrangians where   the conditions $\det ||\{\chi_a, \phi^b\} || \neq 0$
must be satisfied for the Poisson brackets.

The corresponding path integral for the S-matrix in our case is
\be
\langle {\rm out} | S | {\rm in} \rangle= \int \exp \Big ( {i\over h} \int _{-\infty}^{+\infty} (p_i \dot q^i- H(q,p) )dt\Big)\prod _t \delta \Big (\phi^a(q,p)\Big ) \prod_i dp_i (t)dq^i (t)
\ee
In details the canonical momenta conjugate to the $A^I_i, \, i=1,2,3$, are given by
\begin{equation}
\pi^i_I = \frac{\d \mathcal{L}}{\d \dot{A}^I_i} = \mathcal{I}_{IJ}\, (F^J{}_0)^i -\frac{1}{2} \mathcal{R}_{IJ}\,\varepsilon^{ijk}F^J_{jk}, 
\end{equation}
and there is a constraint $\pi^0_I=0$. This relation can be inverted to get
\begin{equation}
\dot{A}^{Ii} = (\mathcal{I}^{-1})^{IJ} \,\pi^i_J + \partial^i A^I_0 + \frac{1}{2} (\mathcal{I}^{-1} \mathcal{R})^I{}_J \,\varepsilon^{ijk} F^J_{jk} ,
\end{equation}
The first-order action of the type given in eq. \rf{Fad} is
\begin{equation}
S (\pi, A, A_0) =  \int\!d^4x\, \left( \pi^i_I \dot{A}^I_i -  H(\pi^i, A_i) -  A^I_0 \, \mathcal{G}_I \right),
\label{1st}\end{equation}
where
\begin{align}
 H (\pi, A) &= \frac{1}{2} (\mathcal{I}^{-1})^{IJ} \pi^i_I \pi_{J i} + \frac{1}{4} (\mathcal{I} + \mathcal{R}\mathcal{I}^{-1}\mathcal{R})_{IJ} F^I_{ij} F^{Jij} + \frac{1}{2} (\mathcal{I}^{-1} \mathcal{R})^I{}_J \,\varepsilon^{ijk} \pi_{Ii} F^J_{jk} \\
\mathcal{G}_I &= - \partial_i \pi^i_I .
\end{align}
 The Lagrange multiplier for the constraint
$
\partial_i \pi^i_I = 0 
$
is a time component of the vector $A^I_0$, which enters the Lagrangian \rf{vec} without a time derivative due to Abelian gauge symmetry. An assumption made in \cite{Henneaux:2017kbx} is that there is no gravity. This helps to proceed with the Hamiltonian analysis in a simple way. Namely, 
in a flat space the constraint $\partial_i \pi^i_I = 0 $  can be  resolved. 
In the flat space 
\begin{equation}
\pi^i_I = -  \varepsilon^{ijk} \partial_j Z_{I k},
\end{equation}
Here  $Z_{I}$ is defined up to  a gauge transformation $Z_{I i} \rightarrow Z_{I i} + \partial_i \tilde{\epsilon}_I$. 
In this way, the remaining Hamiltonian depends only on coordinates $A_i^I=q_i^I$ and on canonical momenta $p^i_I=\pi_I^i$. 

The  action takes the form \cite{Henneaux:2017kbx}
\begin{equation} \label{HJ}
S (\pi^i, A_i)= \frac12 \int \!d^4x \left(  \Omega_{MN} \mathcal{B}^{Mi} \dot{\mathcal{A}}^N_i -   \mathcal{M}_{MN}(\phi) \mathcal{B}^M_i \mathcal{B}^{Ni} \right),
\end{equation}
where the doubled potentials are packed into  3D  vectors\footnote{To avoid confusion with notations used in Sec. 3, we use here the space-time index $i=1,2,3$ for the doublet $\mathcal{A}^M_i $ and its components $A^I_i$ and $Z_{I i}$.}
\begin{equation}
\mathcal{A}^M_i = \begin{pmatrix} A^I_i \\ Z_{I i} \end{pmatrix}, \quad M= 1, \dots, 2n_v,
\label{cA}\end{equation}
and 
\begin{equation}
\mathcal{B}^{Mi} =  \varepsilon^{ijk} \partial_j \mathcal{A}^M_k .
\end{equation}
The matrices $\Omega$ and $\mathcal{M}(\phi)$ are the $2n_v \times 2n_v$ symplectic matrices
\begin{eqnarray}\label{56bein}
\Omega_{MN} = \left(\begin{array}{cc}0 & {\bf 1} \\-{\bf 1} & 0\end{array}\right)  \qquad {\cal M} =\left(
                                        \begin{array}{cc}
                                         \mathcal{I} + \mathcal{R}\mathcal{I}^{-1}\mathcal{R} &\, \, \,  - \mathcal{R} \mathcal{I}^{-1}  \\
     
                                        - \mathcal{I}^{-1} \mathcal{R} &\, \, \,    \mathcal{I}^{-1} \\
                                        \end{array}
                                      \right)                                 
                                      \end{eqnarray}                                     
 Here  E is a symplectic matrix in the fundamental representation of $Sp(2n_v,\mathbb{R})$.                                    In such case, the expression for the S-matrix is given by the path integral of the form, 
 \be
\langle {\rm out} | S | {\rm in} \rangle= \int \exp \Big ( {i\over h} \int d^4x  (P_i ^T \dot Q^i- P^T {\cal M} P) )\Big)\prod _t \prod_x \delta(P^i-P^i(Q))dP_i (x)dQ^i (x)
\label{S}\ee
and
\be
Q_i^M= \mathcal{A}^M_i \, ,\qquad P^i_M = \Omega_{MN}\mathcal{B}^{N i}\, , \qquad 
P^i_M(Q)= \Omega_{MN} \varepsilon^{ijk} \partial_j \mathcal{A}^N_k 
\ee
Now both the canonical coordinates $Q_i^M=\mathcal{A}^M_i$ as well as canonical momenta $P^i_M = \Omega_{MN}\mathcal{B}^{N i} $ are doublets transforming under $Sp(2n_v,\mathbb{R})$
\be
\mathcal{A} = E \mathcal{A}'\, , \qquad \mathcal{B} = E \mathcal{B}' \ ,
\ee
where E is a symplectic matrix in the fundamental representation of $Sp(2n_v,\mathbb{R})$ and 
\be
Q= EQ' \, , \qquad P= (E^T)^{-1} P' \, , \qquad P^T= (P')^T E^{-1}  \ .
\ee  
This means that under  symplectic transformation with account of 
\be
 \Omega' =  \Omega= E^T \Omega E  \, , \qquad {\cM}'(\phi)= E^T \cM(\phi)  E
\ee   
we can change the symplectic frame by preserving the same choice of the coset representative
We find that 
 \be
S(P, Q, \phi)= \int d^4x (P^T \dot Q - P^T {\cal M}(\phi) P)= \int d^4x ((P')^T \dot Q' - (P')^T {\cal M}(\phi) P')= S(P', Q', \phi)
\ee
\be
\delta(P-P(Q))(x)  = (E^T)^{-1} \delta(P'(x)-P'(Q')(x) 
\ee
Finally, the measure of integration 
$dP dQ=  dP' dQ'$ 
is also preserved, up to local terms of the form 
\be
\delta^4(0)
\ee 
since the change of variables is local. The role of these terms was investigated, starting with   \cite{Fradkin:1973wke} and later in \cite{Fradkin:1977hw,Henneaux:1985kr,Grigorian:1992sx}. It was explained in \cite{Fradkin:1973wke} that the relevant terms in the local measure of integration, in general, have to be kept with the purpose to 
 cancel all the divergences of the type $\delta^4(0)$  which arise in the theory in loop integrals. However, already at that time it was noticed that in gauge invariant dimensional regularization, which appeared just a bit earlier in \cite {tHooft:1972tcz}, the relevant terms proportional to $\delta^4(0)$ in the loop Feynman diagrams and in the measure of integration vanish separately, there is no need to keep them for the cancellation purpose. There is an agreement on this at present, and the local measure of integration can be ignored if proper regularization is applied.

We conclude that the $Sp(2n_v,\mathbb{R})$ transformation on the path integral \rf{S} is a canonical change of variables in the path integral
\be
(Q,P) \rightarrow (Q',P')
\ee
that preserves the form of Hamilton equations of motion. As different from the Lagrangian form the Hamiltonian 
formulation  does not require field equations to prove $Sp(2n_v,\mathbb{R})$ symmetry.

To compare supergravities in different symplectic frames, we have to make a choice of the matrix E to belong to a double quotient $G_U (\mathbb{R})  \backslash Sp(2n_v; \mathbb{R})/GL(2n_v)$, as explained around eq. \rf{DQ}.

To change a coset representative $\phi$ into a different one, we call it $\bar \phi$, following \cite{Henneaux:2017kbx},  without changing  symplectic frame we have to take the matrix E in the electric subgroup  of $Sp(2n_v,\mathbb{R})$. Namely
\be
Q\rightarrow \bar Q= F  Q\, , \qquad P\rightarrow \bar P =F^T P \, , \qquad
\phi \rightarrow \bar \phi
\ee
Here $F$ is a lower triangular matrix of the form
\be
\left(\begin{array}{cc}M & 0 \\BM & M^{-T}\end{array}\right), \quad M\in GL(n_v), \quad B^T=B
\ee
This will result in the form of the path integral  where the choice of the coset representative can be made different from the original, but the frame is preserved
 \be
S(P, Q, \phi)= S(\bar P, \bar Q, \bar \phi)
\ee
When both versions of canonical $Sp(2n_v,\mathbb{R})$  transformations are applied to the path integral, one finds that the S-matrix is independent on the choice of the symplectic frame as well as the choice of the gauge-fixing local H-symmetry in unitary gauges.
\be
\langle {\rm out} | S | {\rm in} \rangle_{frame, gauge} =  \langle {\rm out} | S | {\rm in} \rangle_{frame', gauge'}
\ee
This construction is rather satisfactory. It is a promotion of the classical analysis in \cite{deWit:2002vt,deWit:2002vz,deWit:2005ub,deWit:2007kvg}, which states that at the classical level 4D supergravities on-shell are the same for different choice of gauges of a local H-symmetry. The same classical equivalence on-shell is stated in \cite{deWit:2002vt,deWit:2002vz,deWit:2005ub,deWit:2007kvg} when comparing supergravities I and supergravities II, which correspond to different symplectic frames as well as different choices of coset representatives  \cite{Kallosh:2024rdr}.

This gives strong evidence that $\cN\geq 5$ 4D supergravities are gauge equivalent and predicts the absence of local H-symmetry and global G-symmetry anomalies. It would be nice to extend the derivation above to the case of full supergravity action when both gravitons and fermions are added. 

Since the theory has manifest off-shell $Sp(2n_v,\mathbb{R})$ invariance,  a bona fide $Sp(2n_v,\mathbb{R})$ Noether current can be constructed.  This is different from the covariant formulation where there is an on-shell Noether-Gaillard-Zumino conserved current we presented in Sec. \ref{Sec:NGZ}.

In  \cite{Hillmann:2009zf,Bossard:2010dq} the bona fide \E\, Noether current was derived in the Hamiltonian formulation where they have not decoupled gravity so that the constraint term $A_0^I G_I$ is still present in the action \rf{1st}. However, on the constraint surface we expect that the bona fide \E\, Noether current in  \cite{Hillmann:2009zf,Bossard:2010dq} can be promoted to the bona fide $Sp(2n_v,\mathbb{R})$ Noether current.

We would like to comment here on {\it symplectic nature of the Hamiltonian dynamics} where the canonical momenta form a doublet
\begin{equation}
 \eta = \begin{pmatrix} Q \\ P \end{pmatrix}\, , \qquad \dot \eta = \Omega {\partial H \over \partial \eta}\, , \qquad \Omega = \left(\begin{array}{cc}0 & {\bf 1} \\-{\bf 1} & 0\end{array}\right)
\label{cA}\end{equation}
and equations of motion have a symplectic covariant form. The enhanced 4D dualities due to GZ electromagnetic symmetries are symplectic symmetries, which suggests a possible reason why enhanced 4D dualities might have a deeper meaning in quantum theory.

\section{Discussion}

All D $>4$ $\cN>4$ supergravities have UV divergences below critical\footnote{In case of $\cN<5$ the critical loop order is more complicated and has to be studied separately.}
loop order {\cite{Kallosh:2023css,Kallosh:2023dpr,Kallosh:2024lsl}
\be
L_{cr}^{^{D}}= {2\cN +n\over D-2}\, , \qquad n\geq 0
\ee  
 This means that the relevant superinvariant counterterms cannot have local H-symmetry and global G-symmetry. Therefore these UV divergences signify local H-symmetry and global G-symmetry anomalies.

This is consistent with the fact that it is not possible to prove that in D $>4$ supergravities, the S-matrix is independent of the choice of the local H-symmetry gauge-fixing condition: symmetric or Iwasawa type. 

For 4D $\cN\geq 5$, no UV divergences below critical loop order $L_{cr}^{^{4D}}= \cN$ have been discovered so far. 
Therefore, there are no indications that local H symmetry has anomalies.

This is in agreement with our  argument\footnote{The simplified path integral construction in Sec. \ref{Sec:8}\, following \cite{Henneaux:2017kbx} is based on approximation where duality neutral fields, gravitational and fermionic, are absent. For the full proof of quantum gauge equivalence one would need to study the path integral for the complete supergravity action.}
that in $\cN\geq 5$  4D supergravities, the  S-matrix is independent of the choice of different versions of supergravities and different choices of gauges. Supergravities of type I and II in symmetric and Iwasawa gauges must give the same results.

Classically, the proof of gauge equivalence of supergravities is based on the existence of different symplectic frames in 4D supergravity \cite{deWit:2002vt,deWit:2002vz,deWit:2005ub,deWit:2007kvg}. The existence of such different frames, with Lagrangians different off shell, was established using the $Sp(2n_v)$ GZ duality symmetry, modulo scalar and vector field redefinitions. We have checked that the corresponding double quotient 
\be
E^{^{4D}}=G_U (\mathbb{R})\backslash Sp(2n_v,\mathbb{R})/GL(n_v , \mathbb{R})
\label{Bdewit}\ee
is nontrivial for all $\cN\geq 5$. This means that the number of generators in GZ duality group $Sp(2n_v,\mathbb{R})$ has to exceed the number of generators in $G_U$ duality group, known as E7-type groups, plus the number of generators in  $GL(n_v , \mathbb{R})$.

The difference between the 4D case and other even dimensions is that in 6D the relevant GZ duality group $SO(5,5)$ has the same number of generators as  $G_U=$ \EE\,, and in 8D the relevant GZ duality group $Sp(2, \mathbb{R})$ has the same number of generators as $SL(2, \mathbb{R})$ part of the $E_{3(3)}$ $G_U$ duality.

There is also no enhancement of duality symmetry in all odd dimensions as GZ duality is available only in even dimensions D=2k, where electric and magnetic forms have the same dimension k and can form duality doublets, required for GZ symmetry. Therefore in D $>4$ supergravities the corresponding double quotient analog of the one in 4D in eq. \rf{dewit} is trivial,
\be
E^{^{D>4}}={\rm I} \ .
\ee
Therefore there are no tools available to connect symmetric and Iwasawa gauges or supergravities I and II  in D $>4$ supergravities.

Future computations will not be able to change the fact that all D $>4$ supergravities have perturbative UV divergences and local H-symmetry anomalies. However, future 4D loop computations may or may not discover UV divergences.

Other explanations of UV finiteness of the 4-loop superamplitude in $\cN=5$ supergravity in 4D were given before in \cite{Kallosh:2018wzz,Gunaydin:2018kdz,Kallosh:2023asd}, but here we have an explanation of currently available computational data in all integer D$\geq 4$ dimensions. 

One more feature here is that we have found the symmetries explaining enhanced cancellations discovered 10 years ago in  \cite{Bern:2014sna}. In the recent review \cite{Bern:2023zkg}, these cancellations were still qualified as {\it ``puzzling enhanced ultraviolet cancellations, for which no symmetry-based understanding currently exists''}.  Here we have found  {\it the enhanced symmetries predicting enhanced cancellation} of the UV divergences in \cite{Bern:2014sna}. This enhanced duality symmetry in 4D $\cN=5$ supergravity is GZ $Sp(20, \mathbb{R})$ electro-magnetic duality, modulo scalar and vector change of field variables. The corresponding  symmetry generators  are given by a double quotient $SU(1,5)(\mathbb{R}) \backslash Sp(20, \mathbb{R}) / GL(10, \mathbb{R})$.

Our current explanation of UV finiteness in 4D and UV divergences in D $>4$ supergravities is based on the difference between Gaillard-Zumino and U-dualities. From this perspective, if the results of future computations in  4D $\cN\geq 5$ supergravities continue to be UV finite, there will be no surprise, we will just have a confirmation of  anomaly-free  local H symmetry  in 4D $\cN\geq 5$ supergravities. 

\

\noindent{\bf {Acknowledgments:}} I am grateful to  D. Freedman, M. Gunaydin, Y.-t. Huang,  A. Linde, R. Minasian, H.~Nicolai, R. Roiban, C. Wen, and  Y. Yamada for stimulating discussions of the UV divergences and anomalies in supergravities.   I am grateful to H. Samtleben and A. Van Proeyen for their collaboration on a project on gauge-fixing local H-symmetries in supergravities, on which the current work is based.
I am grateful to the participants of the Amplitudes 2024 conference for their interest in this work.
It is supported by SITP and by the US National Science Foundation grant PHY-2310429.

\appendix

\section{4D $\cN=4$ supergravity}
The version of $\cN=4$ pure supergravity presented in \cite{Cremmer:1979up} as a truncation from $\cN=8$ has the following coset space
\be
G/H= {SU(1,1)\over U(1)}= {SU(1,1)\times SU(4)\over U(4)} \ .
\ee
It has 6 vectors, $n_v=6$. The double-quotient is non-trivial
\be
E^{^{4D}}_{\cN=4}=G_U (\mathbb{R})\backslash Sp(2n_v,\mathbb{R})/GL(n_v , \mathbb{R})=SU(1,1)\times SU(4)\backslash Sp(12,\mathbb{R})/GL(6 , \mathbb{R}) \ .
\label{dewit}\ee
Its dimension is 24. 

The role of GZ dualities and the corresponding symplectic frames were investigated in 
\cite{DallAgata:2023ahj} where
the full Lagrangian and supersymmetry transformation rules for the gauged 4D $\cN = 4$ supergravity coupled to an arbitrary number of vector multiplets were given. The ungauged theory in \cite{DallAgata:2023ahj} is taken with a particular choice of the coset space representative.

If this would be the whole story of pure $\cN=4$ supergravity with local $U(4)$ symmetry, we would have a problem to explain its  1-loop anomalies  \cite{Marcus:1985yy,Carrasco:2013ypa} and a 4-loop UV divergence \cite{Bern:2013uka}. 
However, the issue of local symmetries of $\cN=4$ supergravity is complicated because  $\cN = 4 $ {\it supergravity is the highest $\cN$ theory for which a local superconformal theory is available }\cite{Bergshoeff:1980is}. 
The  superconformal version of $\cN=4$ Poincar\'e supergravity based on $SU(2,2|4)$ symmetry was presented and studied in \cite{Bergshoeff:1980is,deRoo:1984zyh,deRoo:1985jh}. 
The theory in \cite{DallAgata:2023ahj}  is related to a Poincar\'e gauge of the superconformal supergravity \cite{Bergshoeff:1980is,deRoo:1984zyh,deRoo:1985jh} with gauge-fixed local $U(4)$ symmetry.

Interestingly, the known explicit actions of $\cN=4$ supergravity are of the $SO(4)$ types \cite{Das:1977uy,Cremmer:1977tc} and of the $SU(4)$ type in \cite{Cremmer:1977tt}. All of these Poincar\'e supergravities were derived starting from a superconformal theory in \cite{deRoo:1985jh,Ferrara:2012ui}.

It was discovered in \cite{Bern:2012cd} that the 3-loop $\cN=4$ supergravity is UV finite. A conjecture was proposed in \cite{Ferrara:2012ui} that this 3-loop UV finiteness might be a consequence of an unbroken at the quantum level classical superconformal symmetry. One of the superconformal symmetries is a local dilatation Weyl symmetry which forbids  UV divergences. However, soon after, it was demonstrated that pure $\cN=4$ supergravity is ultraviolet divergent at four loops \cite{Bern:2013uka}. This means that despite the classical theory presented in \cite{Bergshoeff:1980is,deRoo:1984zyh,deRoo:1985jh} has local superconformal symmetry, including Weyl symmetry at the nonlinear level, it is broken in quantum theory. The 4-loop UV divergence is a manifestation of a Weyl anomaly of the theory in  \cite{Bergshoeff:1980is,deRoo:1984zyh,deRoo:1985jh} which has Weyl symmetry classically. This 4-loop UV divergence is also known to be associated with 1-loop  supergravity anomalies \cite{Marcus:1985yy,Carrasco:2013ypa}.

The classical superconformal $\cN=4$ theory has  a formulation with local $SU(4) \times U(1)$ and rigid $SU(1, 1)\times SO(6)$ invariance \cite{Bergshoeff:1980is,deRoo:1984zyh,deRoo:1985jh}. This rigid symmetry is not extended to maximal GZ $Sp(n_v, \mathbb{R})$ duality due to constraints imposed by the superconformal $SU(2,2|4)$ symmetry. There is no relation between theories studied in \cite{DallAgata:2023ahj}, enjoying maximal $Sp(12, \mathbb{R})$ GZ duality and the superconformal ones in \cite{Bergshoeff:1980is,deRoo:1984zyh,deRoo:1985jh}.  Basically, maximal GZ duality is not covering all available Lagrangians with various local symmetries beyond the ones with local H of the relevant G/H coset space Poincar\'e supergravities. It only relates to the ones where the maximal local symmetry is H.

Thus, the problem of $\cN=4$ supergravity, which is absent in $\cN\geq 5$ cases, is the existence the Lagrangians given in \cite{Bergshoeff:1980is,deRoo:1984zyh,deRoo:1985jh} with local superconformal symmetry $SU(2,2|4)$. The local symmetries, in addition to $H=SU(4)\times U(1)$ of standard Poincar\'e supergravities are: local Weyl symmetry, local conformal transformations, and local special supersymmetry.

For $\cN\geq 5$, only linearized action has $SU(2,2|\cN)$ superconformal symmetry, it is broken at the nonlinear level for $\cN\geq 5$. Therefore all local symmetries of $\cN\geq 5$ supergravities are H-symmetries. Therefore maximal GZ dualities relate all available $\cN\geq 5$ supergravities to each other.

\section{ Amplitudes and improved scaling at infinity, general D vs. D = 4}

An interesting observation was made in amplitude studies in \cite{Edison:2019ovj} where gravity loop integrands from the ultraviolet were studied.\footnote{I am grateful to J. J. Carrasco, A. Edison and J. Parra Martinez for bringing this work to my attention.}

 It was shown there that there is a certain property, called improved scaling at infinity, where there is a difference between amplitudes in general D and in D=4.  In particular, certain poles at infinity are absent in 4D (but not in other D), which requires a conspiracy between individual Feynman integrals contributing to the amplitude. The better behaviour of a sum of Feynman diagrams than of the individual diagrams suggests that there is a symmetry responsible for the cancellation of these poles at infinity when the full set of diagrams is taken into account. 

The {\it symmetry in quantum theory} in the form of Ward Identities is always a statement about the full set of Feynman diagrams contributing to a given process. The reason is that symmetry, as the change of variables in the path integral, can only apply to the full set of diagrams and not to individual Feynman diagrams.

 Improved scaling at infinity is defined in \cite{Edison:2019ovj} as follows. The amplitude for the L-loop diagrams is given in the form  ${\cal A}_n^{L-loop} =\int d^Dl_1\dots d^D l_L \, {\cal I}_n^{L-loop}$ where ${\cal I}_n^{L-loop}$ is a loop integrand. The multi-particle unitarity cut is studied for the 4-particle amplitudes in D dimensions, $p_1, p_2 \to p_3, p_4$. A set of $L+1$ on-shell conditions is imposed on the values of $l_k$, $k=1, \dots , L+1$, $\sum_k l_k = -(p_1+p_2)$.  On this surface, the values of $l_k$ are taken to infinity as follows
 \be
l_k \to l_k+ t q_k\, ,   \qquad (l_k\cdot q_k) = q_k^2 =\sum_k q_k=0 \, ,   \qquad (q_i\cdot q_j) =0 \, ,   \qquad {\rm for} \, \, {\rm all} \, \, i,j
 \ee
 when $t\to \infty$. They find that in this case, the loop integrand for maximal supergravity scales as 
 \be
 F_{SUGRA}\sim {1\over t^4}
 \ee
at any $L$. A more close look at general D and D=4 in maximal supergravities leads to a formula
\be
 F_{SUGRA}\sim {\Delta \over t^4} + \cO {1\over t^5}
\ee  
and for L=2,3 they find that 
\be
\Delta= (\rm{Gram} [q_1 q_2 p_1 p_2 p_3])^2\, ,  \qquad \Delta=0 \quad {\rm at} \, \, D=4
\ee
and there is a drop by one power for D=4
\be
 F_{SUGRA}^{D=4} \sim {1 \over t^5}
\ee
 This investigation shows an 
the existence of an unexpected difference between 4D and $D>4$ in the UV regions of the amplitude integrands.

There is an interesting coincidence with observations in our paper that in all $D>4$, there are UV divergences in maximal supergravities, and in D=4, there is a delay of UV divergences. Secondly, duality symmetries in all $D>4$ maximal supergravities are U-dualities. But in D=4 there is an enhanced duality beyond U-duality.

It remains to be seen whether these 3 coincidences will survive future loop computations.

\section{Evanescent Effects}
 A set of relatively recent studies of UV divergences in quantum gravity called ``evanescent effects'' \cite{Bern:2015xsa} reflects the much earlier story about Quantum Inequivalence/Equivalence of Different Field Representations as discussed in \cite{Duff:1980qv} and \cite{Siegel:1980ax,Fradkin:1984ai,Grisaru:1984vk,Roiban:2012gi}.
 
 It is best described in  \cite{Duff:1982yw}, and later in \cite{Meissner:2017qwm,Kallosh:2016xnm}, which states that conformal anomalies in the form $\cA_{conf} \sim a_s E_4= a_s R^* R^*$ depend on the choice 
 of the representation for particles of various spins. For example, for spin 0 particles, one can take scalars or 2-forms and add 3-form fields without physical degrees of freedom, etc, see table 5 in  \cite{Duff:1982yw}. Note that even the contribution of gravitons, gravitinos and vectors depends on the choice of representations since upon quantization there are spin zero ghosts fields. The analogous situation takes place when conformal anomalies are computed in terms of $\cN=1$ superfields which have various representations.
 
It was shown in \cite{Duff:1980qv}  that in all cases for $\cN\geq 3$ supergravity there is a choice of representations of fields, or $\cN=1$ superfields, which leads to cancellation of conformal anomalies $R^* R^*$. However, the chiral anomalies $RR^*$ were studied later in  \cite{Marcus:1985yy}, where it was shown that chiral anomalies are absent at $\cN\geq 5$ supergravities.  See also \cite{Kallosh:2016xnm} where the explanation of cancellations of conformal/chiral anomalies was given using supersymmetry and dimension of linearized chiral superfields.

The difference between  the choices of supergravities with various representations of fields or superfields and choices of supergravities made in the context of 
enhanced dualities discussed in this paper is of the following nature. It is best explained using the relation between two different $\cN=4$ supergravities in \cite{Nicolai:1980td}. One is the theory in \cite{Cremmer:1977tt}, which has one scalar $\phi$ and one pseudoscalar field B, which appear in the Lagrangian only through $\partial_\mu B$, so the first Lagrangian is $\cL^1 ( \partial_\mu B) $. To get from the theory in \cite{Nicolai:1980td} a Lagrangian with the vanishing trace anomaly, the following procedure is performed. Every appearance of $\partial_\mu B$ in the Lagrangian in \cite{Cremmer:1977tt} is replaced by a vector ${1\over 2} L_\mu$, and a constraint term is added so that the next intermediate  Lagrangian is
 \be
\cL' ( L_\mu, M_{\mu\nu})   =      \cL^1 ( L_\mu )+ {1\over 4} \epsilon ^{\mu\nu\rho\sigma} M_{\mu\nu} \partial_\rho L_\sigma \ .
 \ee
 The final Lagrangian is obtained when algebraic equation of motion allows to eliminate $L_\mu$ and it depends only on $M_{\mu\nu}$. In this way, one can get instead of the first Lagrangian a second one which depends on $M_{\mu\nu}$ instead of B
 \be
 \cL^1 ( \partial_\mu B) \quad \to \quad  \cL^2 (M_{\mu\nu}) \ .
 \label{NT}\ee
 Classically, these two theories are equivalent. However, the trace anomaly vanishes only in the second theory.
 
 Enhanced dualities in 4D supergravity (for example in $\cN=8$) are different. One finds various versions of supergravity by performing an $E_{7(7)}(\mathbb{R}) \backslash Sp(56, \mathbb{R}) / GL(28, \mathbb{R})$ transformation acting on scalars and on vector field strengths which changes the Lagrangian but classically describes the same theory \cite{deWit:2002vt,deWit:2002vz,deWit:2005ub,deWit:2007kvg}. The procedure of changing the representation of the pseudoscalar described above, resulting in eq. \rf{NT},  is not based on a symmetry. Therefore it is not surprising that the conformal anomalies are different.
 
 In the case of enhanced duality, we use the Sp(56) symmetry when constructing the path integral. In this way, the classical equivalence established in \cite{deWit:2002vt,deWit:2002vz,deWit:2005ub,deWit:2007kvg} becomes quantum equivalence, as we argue in Sec. \ref{Sec:8}.
 
 In the case of gravity interacting with scalars in \cite{Bern:2015xsa}, the explicit computation of the 2-loop 4-point amplitude has shown that the $R^3$ UV divergence computed in supergravity by Goroff and Sagnotti changed in the presence of $n_3$ 3-forms, so that 
 \be
 \cL_{R^3} \sim {1\over \epsilon} \Big({209\over 24}- {15\over 2} n_3\Big) \ ,
 \ee
but the finite $\ln \mu^2$ term is not affected by  $n_3$ 3-forms added to the action, or by $n_2$  scalars converted to 2-forms. It is interesting that there is no integer choice of the 3-forms $n_3$ that would eliminate the $R^3$ UV divergence in pure gravity. Therefore, the Einstein-Hilbert action has to be deformed by $R^3$ and by higher derivative terms to describe the BRST invariant theory, see the discussion of this point in \cite{Abreu:2020lyk,Kallosh:2023asd}.

The issue of deformation of extended supergravities required to absorb UV divergences was investigated in \cite{Kallosh:2018mlw,Kallosh:2018wzz,Gunaydin:2018kdz}. The results obtained in \cite{Kallosh:2018mlw}  were inconclusive, but the ones in \cite{Kallosh:2018wzz,Gunaydin:2018kdz} suggested that such deformations is  are inconsistent. It was found more recently in \cite{Kallosh:2023asd} that such a deformation for $\cN\geq 5$ breaks non-linear supersymmetry. A related conclusion was reached in  \cite{Nicolai:2024hqh}  where it was stated that 
`not a single counterterm is known that would be compatible with full non-linear supersymmetry and non-linear \E\,  invariance'.
We believe, therefore, that deformed $\cN\geq 5$ supergravities are inconsistent.

To conclude, the issue of Quantum Inequivalence/Equivalence of Different Field Representations or dual field theories was studied in \cite{Duff:1980qv,Siegel:1980ax,Fradkin:1984ai,Grisaru:1984vk,Roiban:2012gi}, mostly in 4D  and 2D  for specific cases. In our paper, we studied what we have called enhanced dualities in supergravities: they are present in 4D but absent in $D > 4$. That is why we concluded that there is quantum equivalence in 4D at $\cN\geq 5$, but not in $D>4$ in maximal extended supergravities.
 
\bibliographystyle{JHEP}
\bibliography{refs}

\end{document}